\documentclass[12pt, a4paper]{article}
\usepackage{graphicx}
\usepackage{amssymb}
\usepackage{amsmath}
\usepackage{bm}
\usepackage{color}
\usepackage{theorem}
\usepackage{subfigure}

\usepackage[sort&compress,numbers, merge]{natbib}

\definecolor{Orange}{cmyk}{0,0.61,0.87,0}
\definecolor{JungleGreen}{cmyk}{0.99,0,0.52,0}
\definecolor{OliveGreen}{cmyk}{0.64,0,0.95,0.40}
\definecolor{Brown}{cmyk}{0,0.81,1,0.60}
\definecolor{RoyalBlue}{cmyk}{0.71,0.53,0,0.12}

\usepackage[colorlinks=true, linkcolor=OliveGreen, citecolor=RoyalBlue,
urlcolor=Brown]{hyperref}

\begin{document}

\begin{titlepage}

\begin{flushright}
{\tt
KIAS-P17024 \\
UMN-TH-3624/17 \\
UT-17-13
}
\end{flushright}

\vskip 1.35cm
\begin{center}

{\Large
{\bf
Low-Scale $D$-term Inflation 
and the Relaxion
}
}

\vskip 1.5cm

Jason L. Evans$^a$,
Tony Gherghetta$^b$,
Natsumi Nagata$^c$,
and
Marco Peloso$^b$

\vskip 0.8cm

{\it $^a$School of Physics, KIAS, Seoul 130-722, Korea} \\[3pt]
{\it $^b$School of Physics \& Astronomy, University of Minnesota,
 Minneapolis, MN 55455, USA}\\[3pt]
{\it $^c$Department of Physics, University of Tokyo, Tokyo
 113--0033, Japan}

\date{\today}

\vskip 1.5cm

\begin{abstract}
 We present a dynamical cosmological solution that simultaneously
 accounts for the early inflationary stage of the Universe and solves the
 supersymmetric little hierarchy problem via the relaxion mechanism. First,
 we consider an inflationary potential arising from the $D$-term of a new $U(1)$ gauge
 symmetry with a Fayet--Iliopolous term, that is independent of the relaxion.
 A technically natural, small $U(1)$ gauge coupling,  $g\lesssim 10^{-8}$,  allows for a
 low Hubble scale of inflation, $H_I\lesssim 10^5$~GeV, which is shown to be
 consistent with Planck data. This feature is then used to realize a supersymmetric two-field
 relaxion mechanism, where the second field is identified as the
 inflaton provided that $H_I\lesssim 10$~GeV. The inflaton controls the relaxion barrier height allowing
 the relaxion to evolve in the early Universe and scan the
 supersymmetric soft masses. After electroweak symmetry is broken, the
 relaxion settles at a local supersymmetry-breaking minimum with a range
 of $F$-term values that can naturally explain supersymmetric soft mass
 scales up to $10^6$~GeV.

\end{abstract}

\end{center}
\end{titlepage}

\section{Introduction}

A natural solution to the hierarchy problem in the Standard Model (SM)
has motivated the development of particle physics for decades with
predictions of new states near the electroweak scale.  However, to date,
the experimental results at the Large Hadron Collider (LHC) have begun
to call into question of whether naturalness is a relevant guide for
physics beyond the Standard Model. For example, in supersymmetric (SUSY)
models, colored superpartner masses need to be heavier than the TeV
scale to evade LHC searches \cite{ATLAS-CONF-2017-022,
ATLAS-CONF-2017-021, CMS-PAS-SUS-16-036}, exacerbating the tuning, while
constraints on other models addressing naturalness (such as composite Higgs models
\cite{Panico:2015jxa}) lead to a similar conclusion.

Recently a new approach to naturalness, that evades the LHC constraints,
uses the idea of cosmological relaxation \cite{Graham:2015cka} (for
previous studies with a similar idea, see Refs.~\cite{Abbott:1984qf,
Dvali:2003br, Dvali:2004tma}). In this process, an axion-like particle
(the ``relaxion'') associated with a shift symmetry, is coupled directly
to the Higgs field during a nearly de-Sitter phase of the Universe. This
coupling contributes to the mass-squared of the Higgs field, since
initially the relaxion has a very large field value.  During the
cosmological evolution of the relaxion, caused by an explicit breaking of
the shift symmetry, the field value changes and the Higgs mass-squared
is reduced. Eventually the Higgs mass-squared reaches a critical value
where it flips sign, triggering electroweak symmetry breaking with the
Higgs field developing a vacuum expectation value (VEV). The generation
of the Higgs VEV then back reacts on the relaxion potential, causing the
relaxion to stop at a local minimum. The slope of the relaxion
potential, which is proportional to the shift-symmetry breaking
parameter, can then be chosen so that the Higgs VEV is naturally set to
be at the weak scale. This provides a technically natural solution to
the hierarchy problem.

However, the relaxion process itself is not a completely satisfactory
explanation of the hierarchy problem.  First, it can only address
radiative corrections to the Higgs mass that depend on a cutoff scale
which is generally much lower than the Planck scale \cite{Graham:2015cka}.
The hierarchy problem is therefore only partly alleviated. Second, it requires
a very low inflation scale in order to naturally realize a weak scale Higgs
VEV.  However low-scale inflation models often introduce some new tuning
which is not solved by the relaxion process \cite{DiChiara:2015euo}.
The first problem is not a concern if the relaxion process is instead
embedded into a supersymmetric framework.  Supersymmetric models with
soft mass scales near the PeV scale ($10^6$ GeV) are meso-tuned, but can
easily accommodate a 125 GeV Higgs boson mass \cite{Hall:2011jd,
Hall:2012zp, Ibe:2011aa, Ibe:2012hu, Arvanitaki:2012ps,
ArkaniHamed:2012gw, Evans:2013lpa, Evans:2013dza}. Thus if the relaxion
actually scans the supersymmetric soft masses, then the cutoff scale in
the relaxion mechanism can be identified with the soft mass scale. This
naturally provides an explanation for the tuning using just the relaxion
\cite{Batell:2015fma}, or else a two-field relaxion mechanism
\cite{Espinosa:2015eda} can be generalized to supersymmetry
\cite{Evans:2016htp}, which preserves the QCD axion solution to the
strong CP problem. For other recent studies on the relaxion mechanism, see
Refs.~\cite{Hardy:2015laa, Huang:2016dhp, Kobayashi:2016bue,
Hook:2016mqo, Higaki:2016cqb, Choi:2016luu, Flacke:2016szy,
McAllister:2016vzi, Choi:2016kke, Lalak:2016mbv, You:2017kah}.

In this paper we address the second problem in the context of $D$-term
inflation \cite{Stewart:1994ts, Binetruy:1996xj, Halyo:1996pp,
Binetruy:2004hh} and identify the second field in the two-field relaxion
mechanism with the inflaton. In order to realize an inflationary model
with a very low Hubble
scale~\cite{German:2001tz}, the potential must be very flat or else the
density perturbations will not satisfy the cosmic microwave background
(CMB) constraints \cite{Mukhanov:1990me}. One way of accomplishing this
is to take small-field inflation and tune the initial condition so that
the potential is very flat. Although this may work, it is not a very
appealing approach to inflation since it is difficult to justify why the
initial value of the field is so tuned. Furthermore, when applied to
relaxion models, this tuning destroys the naturalness of the relaxion
process and the tuning of the Higgs sector has merely been transferred
to the inflationary sector.

Large field inflation, on the other hand, is in general fairly insensitive to the
initial field value. However, it is difficult to naturally obtain a sufficiently
flat potential to realize the density perturbations if the scale of
inflation is too low.  A model that combines low scale inflation with
the insensitivity to initial conditions, typical of
large field inflation, can be found in supersymmetry.  In $D$-term
inflation~\cite{Stewart:1994ts, Binetruy:1996xj, Halyo:1996pp,
Binetruy:2004hh}, the Fayet--Iliopoulos (FI) term of some new $U(1)$ gauge
symmetry is responsible for inflation, and therefore at tree-level the
potential is completely flat. Although this flatness is broken at the
loop-level, it will provide the right conditions to obtain
low-scale inflation. We will show that a successful model of $D$-term inflation occurs
for a $U(1)$ gauge coupling, $g\simeq 7 \times 10^{-9}$, corresponding to a Hubble scale
$H_I \simeq 10^5$ GeV. Such a low value of $g$ is technically natural since radiative corrections
vanish in the limit $g \to 0$.
With this low Hubble scale, CMB modes are produced approximately 39 $e$-folds
before the end of inflation (contrary to the 50--60 $e$-folds required
in typical models). Within $D$-term inflation this produces a spectral
tilt in agreement with observations. However after inflation ends, there
is the possibility that cosmic strings will
form because the $U(1)$ phase has different values across different
patches in the sky. This problem can be evaded if one considers a
dynamical generation of the FI term
\cite{Domcke:2014zqa}, where the $U(1)$ gauge symmetry is explicitly broken during
inflation, by an amount that negligibly affects the inflationary evolution. This is achieved
via a superpotential coupling which induces a spatial
alignment of the  phase of the $U(1)$ breaking field that prevents the
formation of topological defects at the
end of inflation. Finally, we show that the fields responsible for this
explicit breaking, together with two additional $U(1)$ singlets,
allow for a sufficiently fast conversion of the inflationary energy to
Standard Model fields (a decay through the $D$-term potential is not
fast enough, due to the smallness of the gauge coupling $g$).

With a naturally flat, low-scale inflation model, the inflaton in
$D$-term inflation can now be identified with the second field (the
``amplitudon'') of the supersymmetric two-field relaxion
model~\cite{Evans:2016htp}. In this model the inflaton is coupled to the
relaxion and controls the barrier height of the relaxion potential, and also
helps to avoid a potential isocurvature problem in the original two-field relaxion
model. The slow roll evolution of the inflaton periodically eliminates the relaxion
barrier, allowing the relaxion to move in a step-wise fashion until,
after electroweak symmetry is broken, it is eventually trapped at a
local supersymmetry-breaking minimum.  A quadratic potential with
shift-symmetry breaking mass parameter, $m_S$ controls the slope of the
relaxion potential. For a soft mass scale, $m_{\rm SUSY}=10^5$ GeV this
parameter is constrained to be $10^{-9}\lesssim m_S \lesssim 10^{-6}$
GeV, provided that the Hubble scale satisfies $H_I\lesssim 10$ GeV. This leads to a
model that simultaneously incorporates low-scale $D$-term inflation consistent with
Planck data, and solves the little hierarchy problem in supersymmetric
models, while preserving the QCD axion solution to the strong CP problem.

The plan of the paper is as follows. In Section \ref{sec:dterm} we present a
phenomenologically viable $D$-term inflation model with a low Hubble scale that
is consistent with the CMB data,  does not form cosmic strings, and has a successful
reheating to Standard Model fields. This low-scale inflation model is then combined
with the supersymmetric relaxion mechanism in Section \ref{sec:relaxion}. In Section
\ref{sec:conclusions}, we summarize our results and provide some
concluding remarks. The paper ends with two Appendices that contain
further details of our model. In Appendix \ref{app:dynamicaldterm} we
present the details of the dynamical generation of the $D$-term, which
allows for the explicit breaking of the $U(1)$ symmetry, preventing the formation of cosmic strings,
and helps to facilitate reheating. Other details concerning the quantum and
thermal corrections arising  from the $U(1)$ symmetry breaking are then
discussed in Appendix  \ref{app:inflation}.

\section{$D$-term Inflation}
\label{sec:dterm}


\subsection{$D$-term inflation model}
\label{sec:dtermmodel}

Let us begin by reviewing the $D$-term inflation model
\cite{Stewart:1994ts, Binetruy:1996xj, Halyo:1996pp, Binetruy:2004hh} in
light of recent cosmological observations. The basic model of $D$-term
inflation contains three chiral superfields, $T$, $\Phi_+$, and $\Phi_-$
which have charges of $0$, $+1$, and $-1$ under a $U(1)$ gauge
symmetry, respectively.
This model takes advantage of the FI term of the $U(1)$ gauge symmetry,
with which the auxiliary field of the $U(1)$ gauge field is given by
\begin{equation}
D=g\left(\left|\phi_+\right|^2-\left|\phi_-\right|^2 -\xi \right)~,
\label{eq:Dterm}
\end{equation}
where $g$ is the $U(1)$ gauge coupling, $\xi$ is the FI term, and
$\phi_\pm$ are the scalar components of $\Phi_\pm$. We take $\xi > 0$ in
what follows. The superpotential for this model is
\begin{equation}
W=\kappa \,T \Phi_+\Phi_-~,
\label{eq:WT}
\end{equation}
where $\kappa$ is a dimensionless parameter, which is taken to be real and
positive. We write the scalar components of $T$ as
\begin{equation}
T= \frac{1}{\sqrt{2}}\left(\tau+i\sigma\right)+\dots~.
\end{equation}
In the following discussion, we regard $\sigma$ as the inflaton and
consider the case where $|\sigma| \gg |\tau|$.\footnote{The imaginary
part is chosen to be the inflaton so that when the relaxion mechanism is
discussed, it can more readily be identified with the amplitudon
\cite{Evans:2016htp}, {\it i.e.}, the second field in the two-field
relaxion scenario \cite{Espinosa:2015eda}.} The
tree-level scalar potential for this model is then
\begin{equation}
V_{\rm tree}= \kappa^2 \left[
\frac{\tau^2 + \sigma^2}{2}
 \left(\left| \phi_-\right|^2 + \left|\phi_+\right|^2\right)
+\left|\phi_+\phi_-\right|^2
\right]
+
\frac{g^2}{2} \left[\left|\phi_+\right|^2-\left|\phi_-\right|^2-\xi\right]^2~.
\end{equation}
This potential has a SUSY-preserving minimum at $\tau = \sigma = \phi_-
= 0$ and $|\phi_+| = \sqrt{\xi}$ with $V_{\rm tree} =0$.

If $\sigma$ has a large field value, however, we can find a local minimum with
$V_{\rm tree} > 0$ at which $\phi_+ = \phi_- = 0$, and the charged fields
$\phi_\pm$ have a mass squared
\begin{equation}
 m_\pm^2 = \frac{\kappa^2 \sigma^2}{2} \mp g^2 \xi ~.
\end{equation}
This local minimum becomes unstable when $|\sigma|$ is below the
critical field value
\begin{equation}
 \sigma_c \equiv \frac{g}{\kappa} \sqrt{2\xi} ~.
\label{eq:sigmac}
\end{equation}
Thus, the initial field value of $\sigma$ must satisfy $\sigma \gg
\sigma_c$. As long as this condition is satisfied, the initial value of
the inflaton field is unimportant, like all large-scale inflation
models. At this local minimum, the tree-level potential for $\sigma$ is
completely flat: $V_{\rm tree} = g^2 \xi^2/2$.

However, in order to realize a phenomenologically acceptable model, the
slope of the inflaton potential must be non-zero. Fortunately, the
correct slope is provided by the  quantum corrections encoded in the
Coleman--Weinberg term~\cite{Coleman:1973jx}. Since we will consider
the case $\sigma \gg \sigma_c$, we can expand the Coleman--Weinberg
potential keeping the leading order term in $g^2\xi/(\kappa^2 \sigma^2) $.
In this limit, we find that the potential is well approximated by
\begin{equation}
 V \simeq \frac{g^2 \xi^2}{2} \biggl[
1 + \frac{g^2}{8\pi^2} \ln \biggl(\frac{\kappa^2 \sigma^2}{2 Q^2}\biggr)
\biggr] ~,
\label{eq:cw}
\end{equation}
where $Q$ is a renormalization scale. Using this potential, the
slow-roll parameters are found to be
\begin{align}
 \epsilon &\equiv \frac{M_P^2}{2V^2} \left(\frac{\partial
 V}{\partial\sigma}\right)^2 \simeq\frac{g^4}{32 \pi^4} \left(
\frac{M_P}{\sigma}\right)^2 ~, \\[3pt]
 \eta &\equiv \frac{M_P^2}{V} \frac{\partial^2 V}{\partial \sigma^2}
\simeq -\frac{g^2}{4\pi^2} \left(
\frac{M_P}{\sigma}\right)^2 ~,
\end{align}
where $M_P = 2.4 \times 10^{18}$~GeV denotes the reduced Planck mass.
At this point, it is already clear that $\epsilon \ll |\eta|$, which we will later see is important
for realizing an acceptable inflation model in this context.

\subsection{CMB constraints on $D$-term inflation}
\label{sec:cmbconstraint}

To determine the value of $\epsilon$ and $\eta$, which are constrained by the
CMB data, we need to obtain the value of $\sigma$ at the time
when the CMB modes left the horizon. This value can be determined in
terms of the number of $e$-folds of inflation that occurred after the
CMB was set,
\begin{equation}
N_{\rm CMB} = \int H_I \, dt=\int^{\sigma_{\rm CMB}}_{\sigma_c} \frac{d\sigma}{M_P\sqrt{2\epsilon}}
 =\frac{2\pi^2}{g^2 M_P^2}\left(\sigma_{\rm CMB}^2 -\sigma_c^2\right)~,
\label{eq:NCMBdef}
\end{equation}
where $H_I$ is the Hubble parameter during inflation, and
$\sigma_{\rm CMB}$ and $\sigma_c$ are the field values
when the CMB was set and inflation ends, respectively.
We find that for $\kappa\ll
10^{-2}$, $\eta$ is suppressed and $n_s-1 \equiv d \ln P /d \ln k$ is too large
(where $P$ is the scalar power spectrum and $k$ is the wavenumber).
We therefore
assume that $\kappa$ is large enough so that $\sigma_{\rm CMB}$ is not
near the critical point $\sigma_c$. In this case,
$\sigma_{\rm CMB} \gg \sigma_c$, and thus the expression
(\ref{eq:NCMBdef}) can be solved to give
\begin{equation}
\sigma_{\rm CMB} \simeq \frac{g M_P}{\pi}\sqrt{\frac{N_{\rm CMB}}{2}}
 ~.\label{eq:SigCMB}
\end{equation}
Notice that if $g\ll 1$, we do not need super-Planckian excursions for
inflation to work. As we will show below, low-scale inflation requires
$g$ to be very small and inflation occurs for field values well below
$M_P$.

Equation \eqref{eq:SigCMB} allows us to determine the slow-roll
parameters in terms of $N_{\rm CMB}$:
\begin{equation}
\epsilon_{\rm CMB}= \frac{g^2}{16\pi^2} \frac{1}{N_{\rm CMB}} ~,
~~~~~~\eta_{\rm CMB}= -\frac{1}{2N_{\rm CMB}}~.\label{eq:eps0}
\end{equation}
Using these expressions, the spectral tilt is then found to be
\begin{eqnarray}
n_s-1= 2\eta_{\rm CMB}-6\epsilon_{\rm CMB} \simeq 2\eta_{\rm CMB} =
 -\frac{1}{N_{\rm CMB}} ~,
\label{eq:spectilt}
\end{eqnarray}
where we can neglect $\epsilon_{\rm CMB}$ since it is loop suppressed
relative to $\eta_{\rm CMB}$. As can be seen from this expression,
$\eta_{\rm CMB}$, and therefore $n_s$, only depends on $N_{\rm
CMB}$.
Typically for large-scale
inflation the number of $e$-folds is $\simeq 50$--60. For this
model, this would give a spectral tilt $n_s \gtrsim 0.98$, which is
already excluded by current Planck results \cite{Ade:2015xua}. However,
since we now consider low-scale inflation, $N_{\rm CMB}$ is
modified. In fact, the number of $e$-folds of inflation after the CMB is
set, is significantly altered if the scale of inflation is much
lower than that assumed in ordinary large-scale inflation models.

Let us determine the number of $e$-foldings after the CMB is set for
low-scale inflation. The number of $e$-foldings $N_e(k)$ which
corresponds to a wave-number $k$ is defined by
\begin{equation}
 e^{N_e (k)} \equiv \frac{a_{\rm end}}{a_k} ~,
\end{equation}
where $a_{\rm end}$ is the value of the scale factor at the end of
inflation, and $a_k \equiv k/H_I$. Then, we obtain \cite{Liddle:1993fq,
Liddle:2003as, Dodelson:2003vq}
\begin{align}
 N_e(k) &= -\ln k +\ln H_I +\ln\left(\frac{a_{\rm end}}{a_{\rm
 reh}}\right) +\ln\left(\frac{a_{\rm reh}}{a_{\rm
 eq}}\right) +\ln\left(\frac{a_{\rm eq}}{a_{0}}\right)
\nonumber \\
&= -\ln k +\ln H_I +\frac{1}{3}\ln\left(\frac{\rho_{\rm reh}}{\rho_{\rm
 end}}\right) +\frac{1}{4}\ln\left(\frac{\rho_{\rm eq}}{\rho_{\rm
 reh}}\right) +\ln\left(\frac{a_{\rm eq}}{a_{0}}\right) ~,
\end{align}
where $\rho_{\rm end}$, $\rho_{\rm reh}$, $\rho_{\rm eq}$ are the energy
densities at the end of inflation, at the end of reheating, and at the
time of matter-radiation equality, respectively; $a_{\rm eq}$ and
$a_{0}$ are the scale factors at the time of matter-radiation equality
and the present Universe, respectively. In this derivation, we have
assumed instantaneous reheating with a sudden transition from matter to
radiation domination (where the matter domination is due to the coherent
oscillations of the field $\phi_+$  before it decays;\footnote{As we discuss later, this field carries
most of the inflationary energy right after inflation.}  changes in the
thermalization during reheating modify the value of $N_e (k)$, as discussed in
\cite{Podolsky:2005bw}). This radiation-dominated Universe persists until the time of
matter-radiation equality. Note that this estimation suffers from
uncertainty that originates from the assumption on the cosmological
history; for instance, we obtain a smaller $N_e (k)$ if there is an additional
matter-dominated period between reheating and Big-Bang Nucleosynthesis.

Now we set $k$ equal to the default pivot scale taken by the Planck
collaboration \cite{Ade:2015xua}, $k = 0.05~{\rm Mpc}^{-1}$.
We then obtain
\begin{equation}
N_{\rm CMB} \equiv
 N_e(k=0.05~{\rm Mpc}^{-1}) \simeq 38.9+ \frac{1}{3}\ln \left(\frac{H_{I}}
 {10^5~{\rm GeV}}\right)+\frac{1}{3}\ln \biggl(\frac{\rho^{1/4}_{\rm reh}}
 {100~{\rm GeV}}\biggr)~,
 \label{eq:Ncmb}
\end{equation}
where $\rho_{\rm reh}$ should be understood as the energy density at the time at which the equation of state of the plasma formed at reheating becomes $w=1/3$. We also note that we can disregard the very
small variation of the Hubble parameter, and thus fix $H_I \simeq \rho_{\rm
end}^{1/2}/(\sqrt{3} M_P)$ at $N_e=N_{\rm CMB}$.

\begin{figure}[t]
  \centering
  \includegraphics[width=.6\textwidth]{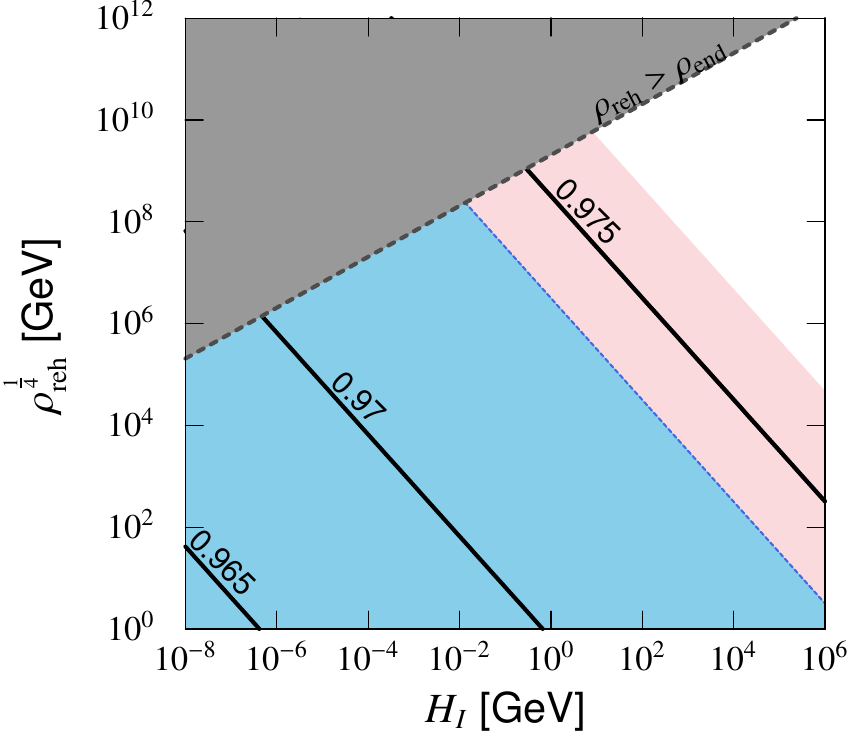}
  \caption{A plot of the reheating temperature (more precisely, $\rho_{\rm reh}^{1/4}$) as a
 function of the Hubble parameter, $H_I$, for various contours of $n_s$
 (0.975, 0.97, and 0.965 from top to bottom, which correspond to
 $N_{\rm CMB} =40$, 33.3, and 28.6, respectively).
 The blue area shows the 1$\sigma$  range given by
 the {\it Planck}$+$BICEP2$+${\it Keck Array} combined results \cite{Array:2015xqh}.
  If one considers only the  {\it Planck} TT$+$lowP result
 \cite{Ade:2015xua}, also the pink area is included at 1$\sigma$.
  The gray shaded region is
 theoretically excluded since $\rho_{\rm reh}$ exceeds the energy
 density of the inflation potential. }
  \label{fig:nsvsHvsRreh}
\end{figure}

Using the expression (\ref{eq:Ncmb}) for the number of $e$-folds after the CMB is set
and the expression for the spectral tilt in Eq.~\eqref{eq:spectilt}, we
show in Fig.~\ref{fig:nsvsHvsRreh} the contours of the spectral tilt as a function of
the reheating energy density and Hubble parameter. The blue area in this figure depicts
the {\it Planck}$+$BICEP2$+${\it Keck Array} combined 1$\sigma$ range for the
spectral tilt  \cite{Array:2015xqh}. The limit obtained from only the {\it Planck} TT$+$lowP
\cite{Ade:2015xua} data, extends the allowed 1$\sigma$ region into the pink area.
The entire parameter space shown in Fig.~\ref{fig:nsvsHvsRreh} falls within the
2$\sigma$ error bands of both results. The gray shaded region is theoretically excluded
since $\rho_{\rm reh}$ exceeds the energy density of the inflation
potential. We thus find that the $D$-term inflation model, with a low Hubble scale
$H_I\lesssim 10^5$~GeV, can actually explain the observed value of $n_s$ with a
sufficiently high ($\gtrsim 100$~GeV) reheating temperature for baryogenesis.

Now that we have seen that an acceptable spectral tilt can be realized
for low-scale inflation, we next need to verify that this model can
generate  cosmological perturbations of the right amplitude. The size of the
cosmological perturbations are determined by the power spectrum which is
related to the inflation scale through
\begin{equation}
 A_s \simeq \frac{V}{24\pi^2 M_{P}^4 \epsilon_{\rm CMB}}
\simeq \frac{\xi^2}{3 (1-n_s) M_{P}^4} ~,
\label{eq:PowSpec}
\end{equation}
where we have used the expression for
$\epsilon_{\rm CMB}$ in Eq.~\eqref{eq:eps0}.
Therefore, the observational value of $A_s$ determines $\sqrt{\xi}$:
\begin{equation}
 \sqrt{\xi} \simeq
9.0\times 10^{15}~{\rm GeV}\times
\left(\frac{1-n_s}{0.03}\right)^{\frac{1}{4}}
\left(\frac{A_s}{2.1\times 10^{-9}}\right)^{\frac{1}{4}} ~.
\label{eq:xisq}
\end{equation}
The gauge coupling $g$ is determined from the Hubble parameter during
inflation via the relation
\begin{equation}
3M_{P}^2 H_{I}^2 \simeq \frac{g^2 \xi^2}{2} ~,
\label{eq:hiandgxi}
\end{equation}
which, using (\ref{eq:xisq}), becomes
\begin{align}
 g \simeq \sqrt{6} \frac{M_{P} H_{I}}{\xi}
 \simeq 7.4\times 10^{-9} \times
\left(\frac{H_{I}}{10^5~{\rm GeV}}\right)
\left(\frac{1-n_s}{0.03}\right)^{-\frac{1}{2}}
\left(\frac{A_s}{2.1\times 10^{-9}}\right)^{-\frac{1}{2}}
~.\label{eq:gu1}
\end{align}
Note that this very small value for $g$ is technically natural since the gauge coupling quantum corrections vanish
in the limit $g\to 0$.  This means that a small gauge coupling at a high energy scale will remain small at low energies, and there is no need to tune the coupling against radiative corrections.

Finally, we study the mass spectrum of this model after the $U(1)$ gauge
symmetry is spontaneously broken. After inflation, $\phi_+$ develops a
VEV of $\langle \phi_+\rangle = \sqrt{\xi} $. This causes $T$ and $\Phi_-$ in the
superpotential \eqref{eq:WT}, to form a vector-like mass term with a mass, $\kappa \sqrt{\xi}$.
Notice that since $\langle \Phi_+ \rangle$ does not
break supersymmetry, the superfield description still holds. As a
consequence, the scalar and fermionic components of $T$ and $\Phi_-$
have an identical mass of $\kappa \sqrt{\xi}$.  On the other hand,
$\Phi_+$ is absorbed by the $U(1)$ gauge vector superfield to form a
massive vector superfield with a mass of
\begin{equation}
m_{Z^\prime} = g \sqrt{2\xi} =9.4\times 10^7~{\rm GeV}\times
\left(\frac{H_{I}}{10^5~{\rm GeV}}\right)
\left(\frac{1-n_s}{0.03}\right)^{-\frac{1}{4}}
\left(\frac{A_s}{2.1\times 10^{-9}}\right)^{-\frac{1}{4}}~.
\label{eq:mzprime}
\end{equation}
More specifically, a massless Nambu--Goldstone boson that originates from
$\phi_+$ is absorbed by the $U(1)$ gauge boson via the gauge
interaction, while the massless fermionic component of $\Phi_+$
combines with the $U(1)$ gaugino via the gaugino interaction to form a
massive Dirac fermion. The radial component of $\phi_+$ acquires a mass,
$g\sqrt{2\xi}$, which is required by supersymmetry to form a massive vector
superfield. As a result, after inflation, we have a vector-like chiral
superfield with a mass of $\kappa \sqrt{\xi}$ and a massive vector
superfield with a mass of $m_{Z^\prime} = g\sqrt{2\xi}$.
\subsection{Cosmic Strings}
\label{sec:cosmic-string}

\subsubsection{Cosmic String Problem}

One complication of $D$-term inflation is the generation of cosmic
strings after inflation ends. When the $U(1)$ symmetry is broken, the
phase of the $U(1)$ breaking field takes different
values in different patches of the sky. This leads to the formation of
cosmic strings \cite{Jeannerot:1997is, Lyth:1997pf}. Since the $U(1)$
symmetry is broken at the end of inflation, these cosmic
strings contribute to the CMB anisotropies, and thus are stringently
constrained by the CMB data \cite{Endo:2003fr, Rocher:2004et, Rocher:2004my,
Wyman:2005tu, Battye:2010hg, Battye:2010xz}. The contribution of cosmic
strings to the CMB angular power spectrum $C_{\ell}^{(\text{str})}$ is
approximately given by
\begin{equation}
 \ell (\ell +1) C_{\ell}^{(\text{str})} = {\cal
O}(100) \times T^2_{\rm CMB} (G\mu)^2 ~,
\end{equation}
where $T_{\rm CMB}$ is the CMB temperature, $G$ is the gravitational
constant, and $\mu$ is the mass per unit length of the string, which is
given by\footnote{As discussed in Sec.~\ref{sec:cmbconstraint}, the
masses of the scalar boson and the $U(1)$ gauge boson
are equal, and thus the cosmic strings that are generated after the $U(1)$
symmetry breaking are Bogomol'nyi--Prasad--Sommerfield (BPS)
strings. }
\begin{equation}
 \mu = 2\pi \langle \phi_+ \rangle^2 = 2\pi \xi ~.
\end{equation}
Therefore, in our model, the size of $G\mu$ is predicted from
Eq.~\eqref{eq:xisq} to be
\begin{equation}
 G\mu \simeq 3.4 \times 10^{-6 } \times
\left(\frac{1-n_s}{0.03}\right)^{\frac{1}{2}}
\left(\frac{A_s}{2.1\times 10^{-9}}\right)^{\frac{1}{2}} ~.
\end{equation}
On the other hand, the Planck 2015 data \cite{Ade:2015xua} gives a
severe bound on this quantity: $ G \mu < 3.3 \times 10^{-7} $.
This clearly shows that the minimal $D$-term inflation model is
disfavored due to the formation of cosmic strings.\footnote{It was previously
argued \cite{Endo:2003fr} that this constraint may be evaded by taking a very small
$\kappa$. In this case, inflation occurs in the vicinity of the critical
value $\sigma_c$ given in Eq.~\eqref{eq:sigmac}, and thus $\sigma_{\rm
CMB} \simeq \sigma_c$. On the other hand,
Eq.~\eqref{eq:PowSpec} shows that if $n_s$ is very close to one, we can
obtain a sufficiently small $\xi$ to evade the cosmic string bound. Such
a value of $n_s$ can be obtained by taking a very small $\kappa$, which makes
$\sigma_{\rm CMB} \simeq \sigma_c$ very large and thus $|\eta|$ very
small. Nevertheless, this possibility is now excluded by the Planck
result, as it restricts the value of $n_s$ and thus $\xi$ cannot become
sufficiently small. }

There are several proposed ways to solve this problem,
however, it is difficult to implement many of them in the context of
low-scale inflation.
One possible solution \cite{Endo:2003fr} is to assume that the
cosmological fluctuations are due to some curvaton. This mechanism,
however, will not end up working for the relaxion process, since there
is also $F$-term SUSY breaking during inflation which generically gives too
large of a mass to the curvaton. Another possible solution is to take a
non-minimal K\"ahler potential \cite{Rocher:2004et, Rocher:2006nh,
Seto:2005qg}.  In these scenarios, either $n_s$ is too
large~\cite{Rocher:2004et, Rocher:2006nh} or the power spectrum scales
down with $g$ \cite{Seto:2005qg}, and so the low-scale implementation is
ruled out by CMB measurements. Another solution is to consider $D$-term
inflation on the part of the potential below the critical point
\cite{Buchmuller:2014dda}. In this regime, inflation occurs after the
$U(1)$ charged field, $\phi_+$ obtains a VEV and so no cosmic strings can
form.  However, this does not work for low-scale inflation since
$\epsilon$ is much too large. A more elaborate
solution is to supplement the $U(1)$ gauge symmetry by a global $SU(2)$
symmetry so that the vacuum manifold is simply connected. In this case,
instead of topologically stable cosmic strings, semilocal strings are
produced when the symmetry is broken \cite{Urrestilla:2004eh}, which are
in general less dangerous compared with stable strings. It turns out,
however, that CMB measurements can restrict even semilocal strings
\cite{Urrestilla:2007sf}, and in fact this solution is disfavored by the
Planck result \cite{Ade:2015xua}. In addition, the presence of a global
$SU(2)$ symmetry leads to the formation of textures, which are again
severely constrained by the Planck data.

Instead in the next subsection we will present one solution that works for our
parameter choices, and that we will later be able to use  in the context
of the relaxion mechanism.

\subsubsection{Dynamical $D$-terms}
\label{sec:dynamicaldtermsol}

Cosmic strings form because the $U(1)$ gauge symmetry breaks after
inflation is over and each patch of the sky has a different phase for
the $U(1)$ breaking field. This can be remedied by explicitly breaking
the $U(1)$ symmetry before the CMB modes exit the horizon during inflation.
This can occur in models where the $D$-term is dynamically
generated \cite{Domcke:2014zqa}, due to a hidden sector breaking of
the $U(1)$ which generates the FI-term. Note that the breaking of this
symmetry must be sufficiently large, so that Hubble fluctuations do not
restore the symmetry in the inflationary sector. This will be accomplished by adding
a marginal coupling in the superpotential. Dynamically generated
$D$-terms are appealing, since they can more naturally explain a FI-term
much smaller then the Planck scale. If the hidden $U(1)$
breaking is appropriately coupled to the visible sector, it will
prevent string formation. The details of how this works are given in
Appendix~\ref{app:dynamicaldterm}, however, we will summarize
the main features below.

Let us add the following superpotential terms to Eq.~\eqref{eq:WT}:
\begin{equation}
\Delta W =\kappa_+T M_+ \Phi_- +\kappa_- T M_- \Phi_+~,
\end{equation}
where $M_\pm$ are the hidden sector fields, which develop VEVs of
${\cal O}(\sqrt{\xi})$ with $\langle M_+ \rangle \neq \langle M_-
\rangle$ to generate the FI-term dynamically. In order not to deform the
$D$-term potential considerably, we take $|\kappa_\pm|$ to be much
smaller than the $U(1)$ gauge coupling $g$.\footnote{Since the couplings
$\kappa_\pm$ explicitly break the shift symmetry for the $T$ field, a mass term
for the $\sigma$ field will be induced. However, this contribution can be sufficiently
small compared with the Coleman--Weinberg effects \eqref{eq:cw} for
$|\kappa_\pm| \ll g$. See Eq.~\eqref{eq:kappauplim} for more a detailed condition.}
Then, using Eq.~(\ref{eq:vfindymod}), we have a linear term for $\phi_+$ during inflation,
\begin{eqnarray}
V\supset \frac{\sigma^2}{2}
\left(\kappa \kappa_+^* \phi_+ M_+^*+ {\rm h.c.} \right)~.
\end{eqnarray}
Because of this linear term, $\phi_+$ has a non-zero VEV during
inflation, with the minimum of the potential occurring for a particular
phase. Since this is the only minimum of the potential until inflation
ends, $\kappa^2 \sigma^2/2\simeq g^2\xi$, the VEV in all patches of the
sky is driven to the same phase, as shown in
Appendix~\ref{app:effoflin}. This prevents the formation of CMB-size cosmic strings,
provided that  the fluctuations in the phase
direction are sufficiently small. As discussed in Appendix~\ref{app:cosstrourmod}, this
requirement gives an upper bound, (\ref{eq:CMBStrings}) on the Hubble parameter during
inflation which becomes
\begin{align}
 H_I &<
1.4\times 10^{8}~{\rm TeV} \times
\left(\frac{|\kappa_+|}{10^{-12}}\right)
\left(\frac{|M_+|}{10^{16} ~{\rm GeV}}\right)
\left(\frac{\kappa}{10^{-2}}\right)^{-\frac{1}{2}}
\left(\frac{A_s}{2.1\times 10^{-9}}\right)^{-\frac{1}{4}}
\left(\frac{1-n_s}{0.03}\right)^{-\frac{1}{2}}~.
\label{eq:kaplowlim}
\end{align}
Thus for the values of $H_I$ satisfying the CMB constraints (see Figure~\ref{fig:nsvsHvsRreh}), we can
always find a value\footnote{The actual constraint on the
relative size of these couplings can be found in
Eq.~\eqref{eq:kappauplim} and the discussion that follows. As $g$, and
thus $H_I$, increases this constraint becomes weaker, and therefore
CMB-sized cosmic strings can always be prevented by choosing $\kappa_+$
appropriately.} of $|\kappa_+| \ll g$, which satisfies this condition
for $M_+ \simeq \sqrt{\xi}$.



Once inflation ends, the universe is reheated. If the reheat temperature
is large enough, thermal fluctuations could generate strings which are
much smaller than the CMB size.  Although these will not appear in the power
spectrum, they could form stable energy configurations which could
overclose the universe. However for this model, as the end of inflation
nears, the VEV of $\phi_+$ grows.  By the time inflation ends, it is
large enough that the VEV of $\phi_+$ is always larger than the maximum
reheat temperature.  Since the temperature sets the size of the thermal
fluctuations of the VEV, no cosmic strings will form from thermal
fluctuations. See Appendix~\ref{app:thermfluc} for the relevant details.

\subsection{Reheating}

With knowledge of  the mass spectrum of the fields in the inflaton
sector obtained in subsection \ref{sec:cmbconstraint}, we can now discuss
reheating after inflation. In Refs.~\cite{Kolda:1998kc,
Kawasaki:2015cla}, kinetic mixing between the gauge fields associated
with the SM hypercharge and the $U(1)$ symmetry driving inflation was
used to reheat to SM fields. However, since the $U(1)$ gauge coupling is
very small, this kinetic mixing is too small to reheat the SM above
the weak scale.\footnote{We assume that the reheating temperature is above the
electroweak scale to facilitate baryogenesis.}  In
Ref.~\cite{Kawano:2008iy}, inflation is driven by the quadratic part of
the $D$-term potential. Inflation models driven by a mass term are no longer
compatible with experimental results. Furthermore, the
method of reheating used there depends on the gauge coupling $g$, and in our case it
would lead to a reheat temperature lower than the weak scale.

Since our low-scale inflation model is a hybrid inflation model, the
energy after inflation is divided between the inflaton, $\sigma$, and
the radial part of $\phi_+$.
This second component is by far the dominant contribution for $g \ll
\kappa$. We thus neglect the energy associated with the
inflaton oscillations in this analysis.\footnote{The energy stored in the inflaton
can be easily dissipated by adding an inflaton coupling to the right-handed
neutrinos, if necessary. Since the inflaton is quite heavy after
inflation, this decay mode can easily thermalize the remaining energy.}
Therefore, in order to transfer the vacuum
energy of inflation into radiation energy, the field $\phi_+$ needs to
decay to lighter states. Because the mass of $\phi_+$ is much smaller
than the scale of its VEV, it is difficult to find viable decay
modes.\footnote{In this context, we use the VEV to mean
the time-evolving homogeneous background value of the field. Because the
VEV of $\phi_+$ begins small, some vacuum decay of $\phi_+$ would be
possible for couplings larger than $g$. However, this decay channel
would shut off once the VEV becomes large enough, leaving the majority of the
energy left in $\langle \phi_+\rangle $.} In fact, generically decay of
$\phi_+$ to fields which couple with a strength greater than $g$ will
be kinematically forbidden. On the other hand, couplings of $\phi_+$
smaller than $g$ would be kinematically allowed but would give a
reheating temperature smaller than the weak scale.

Non-renormalizable couplings do not help. These non-renormalizable
operators would arise from integrating out some heavier fields. The mass
of these heavier fields would, in general, be large since they would
couple directly to $\phi_+$ which has a large VEV. A non-renormalizable
operator with a mass scale of order the VEV of $\phi_+$ would lead to a
suppression of $(m_{\phi_+}/\langle \phi_+\rangle )^{2(n-4)}$, where
$n$ is the dimension of the operator, in the decay width. Even for
$n=5$, this gives too much suppression to obtain a reheating temperature
above the weak scale. If the field couples weakly to $\phi_+$ it could
lead to a smaller mass scale when the particle is integrated
out. However, since it couples weakly to $\phi_+$ this
non-renormalizable interaction gets additional suppression from the
small coupling it has with $\phi_+$, making it difficult to obtain
 a reheating temperature larger then the weak scale.

The problem persists when we consider couplings in the $D$-term
potential. In the $D$-term, other fields couple with $|\phi_+|^2-\xi$,
and so do not receive a large mass from the VEV of $\phi_+$.
However, these couplings are proportional to  $g$, and, due
to the smallness of this parameter, they lead to a reheating temperature
smaller than the weak scale. Nevertheless, this holds the key to reheating
for our model. If we couple particles to $\phi_+$ in a combination where
the VEV cancels, it is possible to reheat above
the weak scale.

A simple example of this method of solving this rather difficult problem
can be found if we again use the fields, $M_\pm$, which generate the dynamical
$D$-term. Using these fields we can couple $\phi_+$ to a
singlet in the following way
\begin{eqnarray}
\Delta W= \kappa_1 R \Phi_+M_- +\kappa_2 R H_u H_d + m_R R\bar R~,
\end{eqnarray}
where both $R$ and $\bar R$ are singlets and $H_{u,d}$ are the MSSM
Higgs superfields.  These new couplings modify the $F$-term of $\phi_+$
and $M_-$ as can be seen in Appendix
\ref{app:dynamicaldterm}.\footnote{We assume that additional
superpotential couplings among these fields are negligibly small.
This still maintains technical naturalness.} These
effects are small because $R$ is stabilized quite close to the
origin. However, they also give new contributions to the
potential\footnote{During inflation, $R$ has a non-zero VEV. However,
this VEV is less than about a GeV for the parameters we consider. This,
plus the fact that $\kappa_2$ generally will be quite small in order to
prevent overclosing the universe, leads to a very small correction to
the Higgs bilinear mass. This small correction to $\mu$ will have a
negligible effect on the relaxion process.}
\begin{eqnarray}
\Delta V_F= \left| \kappa_1 \phi_+M_- +\kappa_2 H_u H_d + m_R \bar
	     R\right|^2 +\left| m_R R\right|^2\label{eq:DelVF}~.
\end{eqnarray}
The cross terms of the above equation gives an interaction for $\phi_+$ of the form
\begin{eqnarray}
-{\cal L}\supset \kappa_2  \left(\kappa_1 \langle M_- \rangle \phi_+
				    +m_R\bar R\right)H_u^\dagger
H_d^\dagger + {\rm h.c.} \label{eq:Phihh}
\end{eqnarray}
The potential in Eq.~\eqref{eq:DelVF} gives an additional contribution to
the mass of $\phi_+$ plus a mixing mass for $\phi_+$ and $\bar R$. If
$\kappa_1 \langle M_- \rangle, m_R \lesssim g\sqrt{2\xi}$, the mass
eigenstates discuss in Section \ref{sec:cmbconstraint} are fairly
unchanged. In this case, the contribution from $\bar R$ can be removed
from the interaction in Eq.~\eqref{eq:Phihh} since $\phi_+$ and $\bar R$
are approximately orthogonal fields. If either $\kappa_1\langle M_-\rangle$ or
$m_R$ is larger than $g\sqrt{2\xi}$, decays coming from this interaction
become suppressed. For $\kappa_1 \langle M_- \rangle \gtrsim
g\sqrt{2\xi}$, the mass of the lightest mass eigenstate coming from $R$
and $\phi_+$ becomes quite light and so cannot decay to Higgs bosons.
If $m_R$ is large, $R$ decouples and all interactions in the
superpotential become suppressed by $m_R^{-1}$, again leading to
suppression of this decay mode.

Given that $\kappa_1 \langle M_- \rangle \lesssim m_{\phi_+}$, we obtain the constraint  $\kappa_1\lesssim g$.  Although this means $\kappa_1$ is a very small coupling, its smallness is offset by a large mass scale $\langle M_-\rangle $, and therefore the trilinear coupling of $\phi_+$ can be as large as $m_{\phi_+}$.

Since the only state in $H_{u,d}$ that is light enough for $\phi_+$ to decay to is the SM like Higgs boson, this interaction becomes
\begin{eqnarray}
-{\cal L}\supset   \frac{1}{2}\kappa_1\kappa_2 \sin2\beta  \langle M_- \rangle \phi_+ h^2~,
\label{eq:phipint}
\end{eqnarray}
where $h$ is the SM-like Higgs boson, $\langle H_{u,d} \rangle = v_{u,d}$, and $\tan \beta = v_u / v_d$.  The interaction (\ref{eq:phipint}) then gives a decay rate for the radial part of $\phi_+$
\begin{eqnarray}
\Gamma_{\phi_+} = \frac{|\kappa_2|^2}{64\pi} \sin^2(2\beta) \left|\frac{\kappa_1\langle M_-\rangle }{m_{\phi_+}}\right|^2 m_{\phi_+}~.
\end{eqnarray}
Recall that the radial part of $\phi_+$ holds the remaining energy of inflation, and therefore the decay produces the reheating temperature
\begin{eqnarray}
T_R=485~ {\rm GeV} \times \left(\frac{106.75}{g_\rho}\right)^{1/4}\left(\frac{\langle M_-\rangle}{10^{16}~{\rm GeV}}\right)\left(\frac{10^8~{\rm GeV}}{m_{\phi_+}}\right)^{1/2}
\left(\frac{\kappa_1}{10^{-9}}\right)
\left(\frac{\kappa_2}{10^{-8}}\right)
~,
\end{eqnarray}
where $g_\rho$ is the number of relativistic degrees of freedom and we have taken $\sin2\beta=1$.

Now, if the mass of $\phi_+$ is lighter then $2m_h$, $\phi_+$ can no longer decay to Higgs bosons. The
$\phi_+$ mass is also given by (\ref{eq:mzprime}), where
it is clear that if $H_I \lesssim 0.1$ GeV, the
decay mode to Higgs bosons shuts off.  In this case, depending on its mass,
$\phi_+$ decays into $ZZ$, $WW$, $b\bar{b}$, {\it etc}, at tree level via the mixing
with the Higgs boson.
Although these decay modes are suppressed by a small mixing angle, it can
still be sufficiently large to allow a reheat temperature above the weak scale.

If $\widetilde R$, the fermionic component of $R$, is lighter than
the Higgsinos, then $\widetilde R$ could be
produced via the Higgsino decay, in addition to the annihilation of the Higgs
fields. Since $\widetilde{R}$ is stable in this case, it may overclose the
universe. There are two ways this can be
avoided. Since, as we discussed above, $\widetilde R$ can be as heavy as
$\phi_+$, it will have a mass as heavy as that in
Eq. (\ref{eq:mzprime}). Experimental constraints allow a bino mass
which is lighter than this, especially if $H_I$ is pushed beyond the weak
scale to make $\phi_+$, and thus $\widetilde{R}$ as well, be heavy
enough.  In this case, $\widetilde{R}$ can decay into bino through
the Higgsino exchange. In the relaxion model below, this type of
spectrum is only realized for some of the parameter space where the bino
mass can be as light as $10^2$~GeV.

The other way to prevent overclosure of the universe from Higgsinos
decaying to $\widetilde R$ is to suppress the reheat temperature below
the Higgsino mass. In this case, the universe will never produce
Higgsinos and so there would be no $\widetilde R$ produced from Higgsino decays.
If the reheat temperature is larger than the bino mass, $\widetilde R$ could still be
produced from bino decays due to bino-Higgsino mixing.
Since the bino can be produced in processes like $hh\to
\widetilde B \widetilde B$, its production cannot be suppressed if the SM
reheats to a temperature above the bino mass. The simplest way to avoid
these problems is to just reheat below the bino mass which requires
$\kappa_2\lesssim 10^{-9}$ for $H_I=10^5$~GeV.\footnote{Such a small $\kappa_2$ also suppresses
the $hh\to \widetilde{R} \widetilde{R}$ process.}
However, it may be possible to reheat above the bino mass in this
scenario if the decay of the bino to $\widetilde R$ is suppressed so
that it happens after the bino freezes out.  In this case the relic
density of the bino could be suppressed during freeze out by some
process such as coannihilation. Since this will effectively reduce the
number of $\widetilde R$ produced from bino decays, it may be possible
to get a relic density of $\widetilde R$ which does not overclose the
universe and may even be the dark matter candidate.\footnote{This is
only possible when the gravitino is heavier than $\widetilde R$ which is
not always the case.}
In the relaxion model we discuss below, only gauginos can be much
lighter than the SUSY-breaking scale. Thus, candidates for the
coannihilation partner of the bino are the gluino or wino. For the bino-gluino
coannihilation case, the bino abundance falls into a desirable value
if the mass difference between the bino and gluino is $\lesssim 100$~GeV
and squark masses are $\lesssim {\cal O}(100)$~TeV \cite{Ellis:2015vaa,
Nagata:2015hha, Ellis:2015vna}. In the case of the
bino-wino coannihilation, on the other hand, the bino-wino mass
difference should be $\lesssim {\cal O}(10)$~GeV
\cite{Nagata:2015pra}. These coannihilation scenarios may be probed at
the LHC by searching for displaced vertex signals \cite{Nagata:2015hha,
Nagata:2015pra, Rolbiecki:2015gsa}. For detailed discussions on these
coannihilation scenarios, see Refs.~\cite{Ellis:2015vaa, Nagata:2015hha,
Nagata:2015pra, Ellis:2015vna} and references therein. Another option is
to assume that there is a wino with a mass of $\gtrsim 500$~GeV
\cite{moriond17, moriond17conf} or a
gluino\footnote{If gluino is lighter than bino, wino or gravitino needs
to be lighter than the gluino in order to make it decay into these
particles. } with a mass of $\gtrsim 2$~TeV \cite{ATLAS-CONF-2017-022,
ATLAS-CONF-2017-021, CMS-PAS-SUS-16-036}, and the bino is heavier than these
particles. In this case, the bino mainly decays into these particles, while
the abundance of these particles are sufficiently suppressed. A few TeV
gluino can be probed at the LHC in the multi-jets plus missing energy
channel \cite{ATL-PHYS-PUB-2014-010}, while an ${\cal O}(100)$~GeV wino can
be probed in the disappearing-track searches \cite{Ibe:2006de,
Fukuda:2017jmk}. Even in these cases, we need to take $\kappa_2$ to be a
small value to suppress the direct $\widetilde{R}$ production process
$hh\to \widetilde{R} \widetilde{R}$.


\section{The Inflaton as an Amplitudon}
\label{sec:relaxion}

The above $D$-term inflation model provides a technically natural realization of
low-scale inflation that is consistent with the current Planck
results. Since low-scale inflation is needed for the relaxion process
and requires a very flat potential, it suggests that the inflaton of
this $D$-term inflation model can be identified with the second
(amplitudon) field in the supersymmetric two-field relaxion model
discussed in Ref.~\cite{Evans:2016htp}. We next present a
relaxion model that combines these two ideas, thereby relating inflation
with solving the supersymmetric little hierarchy problem. In fact,
regarding the amplitudon as the inflaton is also desirable from the
phenomenological point of view; in the minimal setup discussed in
Ref.~\cite{Evans:2016htp}, the light amplitudon field may cause an
isocurvature problem, but we can evade this once we identify it with the
inflaton.

In the supersymmetric relaxion mechanism, supersymmetry breaking in the
visible sector is determined by the $F$-term of the relaxion
superfield. Because of this, the determinant of the Higgs mass matrix is
dependent on the relaxion field value. Initially, the relaxion field
value is large and the determinant of the Higgs mass matrix is positive.
As the relaxion field rolls, the determinant of the Higgs mass matrix
eventually becomes negative and electroweak symmetry breaking
occurs. Electroweak symmetry breaking generates an additional
contribution to the relaxion potential which stops the relaxion from
rolling. For properly chosen parameters, the relaxion stops in a local
minimum that corresponds to a weak scale Higgs VEV.

\subsection{The Inflaton-Relaxion Model}

In Ref.~\cite{Evans:2016htp}, a two-field relaxion model was considered
with an additional field coined the amplitudon. This field was
responsible for controlling the relaxion barrier height and allowing the
relaxion to roll. If we now identify the inflaton of the previous
section (contained in $T$) with this amplitudon, the superpotential for this scenario
becomes
\begin{eqnarray}
W= \kappa T \Phi_+\Phi_-+ \frac{1}{2} m_TT^2+\frac{1}{2} m_S S^2
+\left( m_N +ig_S S+ig_T T+\frac{\lambda}{M_L}H_u H_d\right) N\bar N~,
\label{eq:Sup}
\end{eqnarray}
where the imaginary scalar component of the superfield $S$ is the
relaxion, $N, {\bar N}$ are superfields charged under a strongly-coupled
gauge group ($SU(N)$) and $H_{u,d}$ are the Higgs superfields. The
couplings $\lambda, \kappa, g_{S,T}$ are dimensionless (where $\kappa$
was already introduced in eq.~(\ref{eq:WT}))  and $m_{N,S,T},
M_L$ are mass parameters. Note that $m_S$ is a shift-symmetry breaking
parameter that causes the relaxion to roll, and $\Phi_\pm$ are again
charged under some additional $U(1)$ so that inflation proceeds as it
did in the previous section. In addition $m_T$ is a shift symmetry breaking parameter
that controls the inflaton evolution during the relaxion epoch.
We also consider an identical $D$-term to
the one in Eq.~\eqref{eq:Dterm}. Comparing this to the model in
Ref.~\cite{Evans:2016htp}, the only difference in the superpotential is
the addition of the coupling of the amplitudon with two scalar fields, $\phi_\pm$.
This is the same interaction that we studied in the previous section for the
inflaton.

In addition to these superpotential interactions, the relaxion
superfield, $S$, is coupled to the gauge kinetic function,
\begin{equation}
{\cal L}\supset \int d^2\theta  \left( \frac{1}{2g_a^2}-i\frac{\Theta_a}{16\pi^2}
-\frac{c_a S}{16\pi^2f_\phi}\right){\rm Tr}({\cal W}_a{\cal W}_a) +{\rm h.c.}~,
\label{eq:GaugKin}
\end{equation}
in a similar way to the QCD axion\footnote{The theta term, $\Theta_a$ can be neglected
since it is subdominant compared to the effective value obtained in the early universe from the
large $(\gg f_\phi)$ field value of $\phi$.}, where $f_\phi$ is the global symmetry
breaking scale, $c_a$ is an order one constant and $a$ runs over the SM gauge symmetries
as well as an additional confining $SU(N)$. When this $SU(N)$ confines, the fermionic
components of $N$ and $\bar N$ condense and generate a
$\cos(\phi/f_\phi)$ potential, which is the back reaction that stops the relaxion.

Writing the scalar field components as $S= \frac{s+i\phi}{\sqrt{2}}$ and
$T=\frac{\tau+i\sigma}{\sqrt{2}}$, the relevant parts of these
superfields for our discussion are the relaxion $\phi$, and the
amplitudon (inflaton), $\sigma$. The relaxion and amplitudon correspond
to the Nambu-Goldstone boson of some broken symmetry, and therefore
transform under a shift symmetry. If these shift symmetries are exact,
the potential for these fields would be completely flat. This flatness
is lifted by the explicit breaking of the shift symmetry\footnote{This
shift symmetry will preserve the very flat potential for the inflaton,
$\sigma$.  Higher-order shift symmetric terms in the K\"ahler potential,
$K = K(S+S^\dagger, T+T^\dagger)$, stabilize $s$ and $\tau$
near the origin.
} due to the couplings $m_S$, $m_T$, and $\kappa$ in
Eq.~\eqref{eq:Sup}. The scalar potential is then
\begin{equation}
V_{\text{explicit}}=\frac{1}{2} |m_S|^2 \phi^2
 +\frac{1}{2} |m_T|^2\sigma^2+ \frac{g^4\xi^2}{16\pi^2} \ln
 \left[\frac{\left|\kappa\right|^2 \sigma^2}{2Q^2}\right]~.
\label{eq:ExBr}
\end{equation}
The explicit breaking of the shift symmetry for the amplitudon arises from the
mass term and from integrating out the $\phi_\pm$, which are heavy during
both the relaxion and inflation epochs. As we will see below, the
shift-symmetry breaking mass terms $|m_S|$ and $|m_T|$ are taken to be
very small; such small shift-symmetry breaking effects may be explained
by means of the ``clockwork'' mechanism \cite{Choi:2015fiu,
Kaplan:2015fuy, Giudice:2016yja}.\footnote{A similar idea was first
considered in the context of inflation model building \cite{Kim:2004rp}. }

\begin{figure}[t]
\centering
\includegraphics[clip, width =  0.75\textwidth]{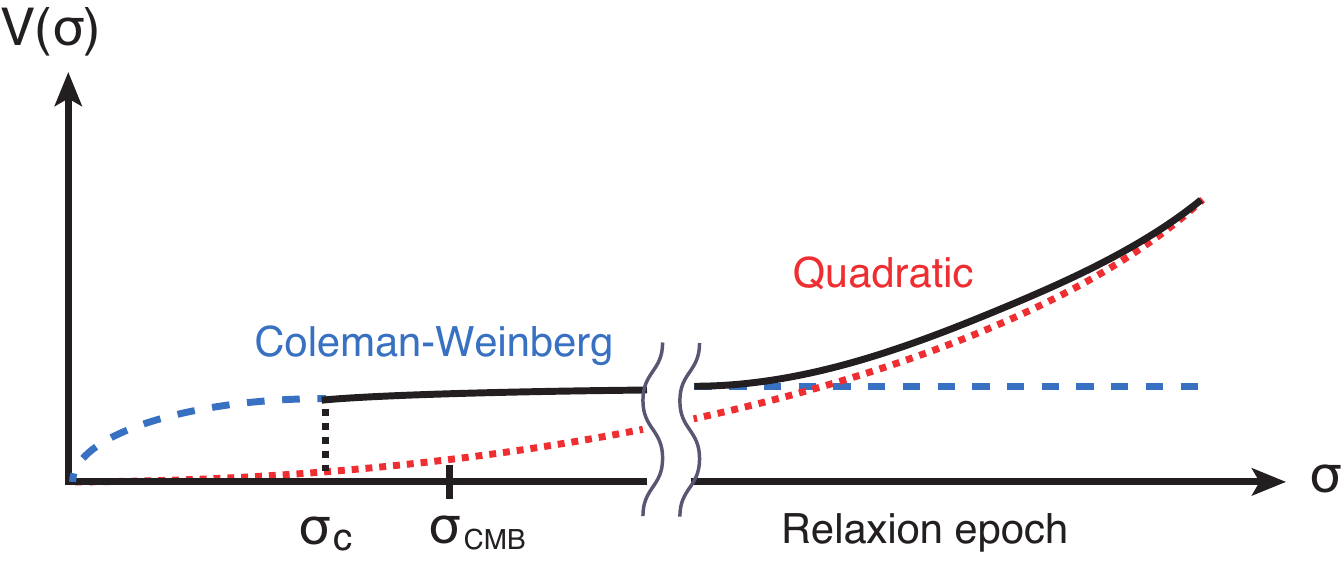}
\caption{Schematic description of the scalar potential as a function of
 the inflaton/amplitudon field, $\sigma$.  The dashed (dotted) lines represent the pure
 Coleman-Weinberg (quadratic) potential, while the solid line is the sum of the two potentials.
\label{fig:inflation}
}
\end{figure}

To combine these two theories, we need the explicit mass term for the
inflaton to dominate during the relaxion epoch and the loop induced
mass to dominate during inflation, as schematically depicted in
Fig.~\ref{fig:inflation}. The ratio of these two masses is
\begin{equation}
R_m= \frac{g^4\xi^2}{8\pi^2\sigma^2} \frac{1}{|m_T|^2} ~.
\end{equation}
During the relaxion epoch the ratio is
\begin{align}
R_m &=\frac{3}{4\pi^2} \frac{1}{A_s(1-n_s)}\frac{H_I^4}{m_{\rm
 SUSY}^2f_\phi^2} \nonumber\\[3pt]
&\simeq 10^{-11}\times
\left(\frac{A_s}{2.1\times 10^{-9}}\right)^{-1}
\left(\frac{1-n_s}{0.03}\right)^{-1}\left(\frac{H_I}{1~{\rm
 GeV}}\right)^4\left(\frac{10^5~{\rm
 GeV}}{m_{\rm SUSY}}\right)^{4}\left(\frac{r_{\rm SUSY}}{1}\right)^{2}
~,
\end{align}
where we have used $m_T^2\sigma^2\sim m_S^2\phi^2 \sim m_{\rm
SUSY}^2f_\phi^2$ which comes from the
constraints\footnote{This relies on the parameterization of the hidden
sector parameters.} on the relaxion
mechanism in Section~\ref{sec:relaxconstraints},
 and Eq.~\eqref{eq:PowSpec} and Eq.~\eqref{eq:gu1}
for $g^4\xi^2$. From this expression, it is clear that $m_T$ dominates
in this regime. Note that we have changed our normalization of $H_I$
in this section since $H_I\sim 10^5$ GeV will no longer be compatible with
the relaxion process (see Eq.~(\ref{eq:HIrelaxbound})).
During the CMB epoch, on the other hand, the ratio is
\begin{equation}
R_m=\frac{3}{4}(1-n_s)\left(\frac{H_I}{|m_T|}\right)^2
\simeq 2 \times 10^{12} \times
 \left(\frac{1-n_s}{0.03}\right)\left(\frac{H_I}{1~{\rm
 GeV}}\right)^2\left(\frac{10^{-7}~{\rm GeV}}{|m_T|}\right)^2 ~,
\end{equation}
where we have used Eq.~\eqref{eq:SigCMB}, or, $\sigma_{\rm CMB} =
H_I/(\pi (1-n_s) A_s^{\frac{1}{2}})$.  For this regime of the potential,
the loop induced mass dominates.  This is just the correct behavior that is
needed to use the $\sigma$ field as both the amplitudon and the
inflaton. Therefore, we can ignore the Coleman-Weinberg
contribution to the mass during the relaxion epoch and the
constraints reduce to those found in Ref.~\cite{Evans:2016htp}, which will
be summarized in Section~\ref{sec:relaxconstraints}.

The back reaction potential for the relaxion is generated by the fields $N,\bar N$ in Eq.~\eqref{eq:Sup}. The $N,\bar N$ fields are charged under the same $SU(N)$ gauge theory that $S$ is coupled to in Eq.~\eqref{eq:GaugKin}.  When the fermionic components of $N,\bar N$ confine at the scale $\Lambda_N$, they give a contribution to the scalar potential of the form
\begin{eqnarray}
&&{\cal A}(\phi,\sigma,H_uH_d)= \left[ m_N -\frac{1}{\sqrt{2}}(g_S\phi + g_T \sigma)
+\frac{\lambda}{M_L} H_u H_d \right]~,\nonumber \\
&& V_{\text{period}} = {\cal A}(\phi,\sigma,H_uH_d) \Lambda_N^3
\cos \left(\frac{\phi}{\sqrt{2}f_\phi}\right) ~,
\label{eq:periodicpot}
\end{eqnarray}
where we have assumed $c_a=1$.
For the model we consider, we take $g_S>0$ and $g_T<0$.  When the $\sigma$ field value is very large, the above potential \eqref{eq:periodicpot} provides a large barrier for the relaxion and therefore the relaxion is initially stabilized at some very large field value. However the inflaton, $\sigma$, is free to roll.  As $\sigma$ rolls, the barrier height is reduced until the mass term in Eq.~\eqref{eq:ExBr} dominates and $\phi$ begins to roll, tracking $\sigma$. This evolution continues until the determinant of the Higgs mass matrix becomes negative and electroweak symmetry is broken. As the Higgs VEV increases, a new barrier develops in the relaxion potential (from the $\lambda$ term in Eq.~\eqref{eq:periodicpot}) eventually stopping the relaxion at a local minimum. The explicit symmetry breaking parameter, $m_S$ is chosen so that this minimum corresponds to a Higgs field with a weak scale VEV.

\subsection{The Constraints on the Cosmological Evolution}
\label{sec:relaxconstraints}

Next we examine the constraints on the cosmological evolution that will limit the parameter space of the relaxion.  To determine these constraints, the relevant part of the scalar potential is given as
\begin{eqnarray}
V= V_{\text{explicit}} + V_{\text{period}}~,
\end{eqnarray}
which are given, respectively, in eq. (\ref{eq:ExBr}) and (\ref{eq:periodicpot}).
However, as we argued above, we can ignore the Coleman-Weinberg
contribution during the relaxion epoch. The main constraints on the
parameter space are as follows:\footnote{For the discussion of other
conditions which lead to weaker constraints, see
Ref.~\cite{Evans:2016htp}.}

\vspace{2mm}
\noindent $\bullet$ {\bf  Inflaton/Amplitudon slow roll:}
In order for the relaxion process to work, we first need the slow roll of the inflaton, $\sigma$, to proceed unimpeded, with little effect from the coupling to the relaxion. The equations of motion for $\sigma$ in the slow roll regime are
\begin{eqnarray}
\frac{d\sigma}{dt} = -\frac{1}{3H_I} \frac{\partial V}{\partial \sigma} =   -\frac{1}{3H_I}\left[m_T^2\sigma -\frac{g_T}{\sqrt{2}}\Lambda_N^3\cos \left(\frac{\phi}{f_\phi}\right)\right]~.
\label{eq:sigEOM}
\end{eqnarray}
Since we need the inflaton rolling to be unaffected by the periodic potential piece in Eq.~\eqref{eq:sigEOM},
we require that
\begin{eqnarray}
m_T^2\sigma \gg \frac{g_T}{\sqrt{2}}\Lambda_N^3~.
\label{eq:Sigma}
\end{eqnarray}
We can remove the $\sigma$ field dependence in this expression by using
the fact that right before electroweak symmetry breaking (EWSB),
$g_S\phi_*\sim - g_T\sigma_*$ (which follows from taking the expression
in the square brackets of Eq.~\eqref{eq:periodicpot} to be zero) and
$\mu_0\sim m_{\rm SUSY}\sim m_S \phi_*/f_\phi$ (which follows from
having a negative determinant of the Higgs mass matrix, see
Ref.~\cite{Evans:2016htp}), with $\phi_*$ and $\sigma_*$ the field
values when the relaxion stops rolling. Using these relationships, the condition
\eqref{eq:Sigma} becomes
\begin{eqnarray}
     \frac{g_T^2}{g_S}\ll \frac{m_{\rm SUSY}f_\phi}{\Lambda_N^3} \frac{|m_T|^2}{|m_S|}~.
     \label{eq:sigSloRol}
\end{eqnarray}


\vspace{2mm}
\noindent $\bullet$ {\bf  Relaxion initial condition:}
Next, we examine the initial condition for the relaxion, $\phi$ which we require to be trapped at a local minimum. This requires that the contribution to the mass of $\phi$ coming from $V_{\text{explicit}}$ is subdominant compared to the contribution coming from $V_{\text{period}}$.  This results in the following constraint
\begin{equation}
|m_S|^2 \ll g_S\frac{\Lambda_N^3}{f_\phi} ~,
\label{eq:phistop2}
\end{equation}
where we have again used the fact that right before EWSB, $g_S \phi_* \sim -g_T\sigma_*$.

\vspace{2mm}
\noindent $\bullet$ {\bf  Stability of relaxion minimum:}
The next constraint we consider comes from requiring that the Higgs VEV does indeed provide a barrier to eventually stop $\phi$ from rolling, and stabilize the relaxion at a local minimum.  This
expression is found from the minimization condition with the following inequality arising from taking $\sin\left(\frac{\phi}{f}\right)=1$
\begin{equation}
 |m_S| \lesssim \frac{|\lambda| \sin 2\beta}{4 M_L} \frac{v^2 \Lambda_N^3}{m_{\text{SUSY}}f_\phi^2}~,
\label{eq:lamovlam}
\end{equation}
where $v=\sqrt{v_u^2+v_d^2}$ is the electroweak VEV. In addition, the term in $V_{\text{period}}$
proportional to $\lambda$ generates a contribution to the soft SUSY
breaking $B_\mu$ term, which causes the determinant of the Higgs mass matrix to oscillate. Requiring that the amplitude of this oscillation be smaller than the electroweak scale gives
the constraint
\begin{eqnarray}
|\lambda|\lesssim \frac{4 M_L v^2}{\Lambda_N^3 \sin 2\beta}~.
\end{eqnarray}
Combining this with Eq.~\eqref{eq:lamovlam}, we find
\begin{eqnarray}
 |m_S| \lesssim
\frac{v^4}{m_{\text{SUSY}}f_\phi^2}~.
\label{eq:StoPhiLam}
\end{eqnarray}


\vspace{2mm}
\noindent $\bullet$ {\bf  Classical rolling condition:}
Another relevant constraint to this scenario comes from requiring that the relaxion, $\phi$, and the inflaton, $\sigma$, undergo classical rolling. The classical rolling conditions are determined from $\dot \sigma / H_I > H_I$, leading to the constraint
\begin{eqnarray}
\frac{|m_T|^2}{|m_S|} \frac{g_S}{|g_T|} m_{\rm SUSY} f_\phi \gg 3 H_I^3~.
\label{eq:ClasRol}
\end{eqnarray}

\vspace{2mm}
\noindent $\bullet$ {\bf $\phi$ tracks $\sigma$ after EWSB:}
In order for $\phi$ to settle in its minimum with the Higgs VEV of order
the weak scale, ${\cal A}(\phi,\sigma,H_uH_d)$ needs to grow quickly
enough with the Higgs VEV so that $\phi$ can stop tracking $\sigma$. By
examining the evolution of ${\cal A}(\phi,\sigma,H_uH_d)$ as the Higgs
VEV develops, we find that as along as\footnote{The additional factors
in Eq.~\eqref{eq:phidecop} as compared to the corresponding expression
in Ref.~\cite{Evans:2016htp} come from considering the contribution to
$B_\mu$ originating in the Higgs dependent part of ${\cal
A}(\phi,\sigma,H_uH_d)$. This oscillatory contribution to $B_\mu$ gives
the dominant contribution to $\phi \frac{d {\cal D}(\phi)}{d \phi}\sim
m_{\rm SUSY}^4\frac{f_\phi^2}{v^2\sin2\beta}$. Following the same
calculation as in Ref.~\cite{Evans:2016htp}, with this single change,
gives the constraint in Eq.~\eqref{eq:phidecop}.}.
\begin{equation}
 \frac{g_S}{\sin2\beta} \frac{m_h^2}{m^2_{\text{SUSY}}}  \frac{\Lambda_N^3}{f_\phi}\frac{v^2}{f_\phi^2}
\lesssim \frac{|m_S|^2}{1-\frac{|m_T|^2}{|m_S|^2}} ~,
\label{eq:phidecop}
\end{equation}
is satisfied, $\phi$ will discontinue its tracking of $\sigma$ with the
Higgs VEV of order the weak scale.

\vspace{2mm}
\noindent $\bullet$ {\bf  Loop corrections to inflaton mass:}
A new constraint, which is only present when the amplitudon is
identified as the inflaton, comes from loop corrections to the mass of
$\sigma$. First, because the coupling $\kappa$ breaks the shift
symmetry, the K\"ahler potential will be affected by this shift-symmetry
breaking at the loop-level,
\begin{eqnarray}
\Delta K \simeq \frac{\kappa^2}{16\pi^2} |T|^2~.
\end{eqnarray}
Since $\kappa$ is the order parameter of this shift-symmetry breaking, it will control the size of all
shift-symmetry breaking in the K\"ahler potential. Because of the inflation constraints discussed above,
this parameter must satisfy, $\kappa \gtrsim 10^{-2}$. If we include this loop-corrected K\"ahler contribution
in the supergravity scalar potential,
\begin{eqnarray}
V_{\rm SUGRA}=e^{\frac{K}{M_P^2}}\left(D_iW K^{i\bar j} D_{\bar j}\bar
				  W-3\frac{|W|^2}{M_P^2}\right)~,
\label{eq:SUGPot}
\end{eqnarray}
we see that there can be important affects on the amplitudon. With the
vacuum energy non-zero during the relaxion process, the exponential
$\exp(K/M_P^2)$, which now depends on $\sigma$ because of the
shift-symmetry breaking, will generate a mass for the inflaton. The
exponential piece can be important because the vacuum energy during the
relaxion process changes at least by an amount of order
\begin{eqnarray}
\Delta V=m_S^2\phi^{2}_*=m_{\rm SUSY}^2f_\phi^2~.
\label{eq:VacEne}
\end{eqnarray}
Expanding the exponential in Eq.~\eqref{eq:SUGPot}, and using Eq.~\eqref{eq:VacEne} for the vacuum energy, we obtain an inflaton mass of order
\begin{equation}
\Delta m_\sigma \simeq \frac{\kappa}{4\pi}\frac{m_{\rm
 SUSY} f_\phi}{M_P}
=  3.3 \times 10^{-12}~{\rm GeV}\times
 \left(\frac{\kappa}{10^{-2}}\right)\left(\frac{m_{\rm
 SUSY}}{10^{5}~{\rm GeV}}\right) \left(\frac{f_\phi}{10^{5}~{\rm
 GeV}}\right)~.
\label{eq:delmphi}
\end{equation}
Now in order for the relaxion process to be viable, this correction to the $\sigma$ mass must be smaller than $m_T$ in the superpotential.

Second, the soft SUSY-breaking effects in the $\phi_\pm$ fields can
induce the $\sigma$ mass term via the Coleman--Weinberg potential of
order
\begin{equation}
 \Delta m_\sigma \simeq \frac{\kappa}{4\pi} \widetilde{m}_{\phi_\pm}
  ,
\label{eq:delmphicw}
\end{equation}
where $\widetilde{m}_{\phi_\pm}$ denote the soft masses of
$\phi_\pm$. If $\widetilde{m}_{\phi_\pm}$ is induced by the Planck
suppressed $\phi_\pm$--relaxion operators, then we expect
$\widetilde{m}_{\phi_\pm} \sim m_{\rm SUSY} f_\phi/M_P$ and thus the
contribution \eqref{eq:delmphicw} is of the same order as
\eqref{eq:delmphi}. If, on the other hand, there is another source of
SUSY-breaking and it gives a larger contribution to
$\widetilde{m}_{\phi_\pm}$, then this gives a more severe constraint, as
we see in Appendix~\ref{app:dynamicaldterm}.

\subsubsection{Combined constraints}

\begin{figure}[t]
\centering
\includegraphics[clip, width =  0.6\textwidth]{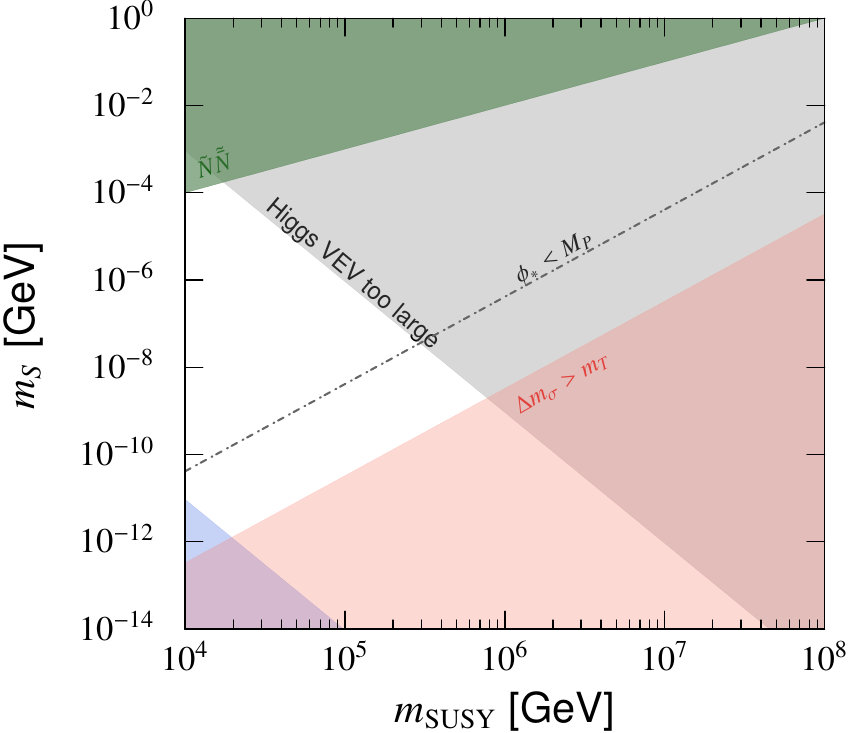}
\caption{{The allowed parameter region in the $m_{\rm SUSY}$--$m_S$
 plane, where $\zeta = 10^{-8}$, $r_{TS}=0.1$, $r_\Lambda = 1$, $r_{\rm
 SUSY} = 1$, and $\kappa = 10^{-2}$.}}
\label{fig8100}
\end{figure}

The relevant constraints can now be combined to restrict the parameter
space of the inflaton-relaxion model. However, to simplify the parameter
space, we will redefine the parameters in a similar manner as was done
in Ref.~\cite{Evans:2016htp}:
\begin{align}
 g_S &= \zeta \frac{m_S}{f_\phi}, ~~~~
 g_S = \zeta \frac{m_T}{f_\sigma}, ~~~~
 f\equiv f_\phi = f_\sigma, ~~~~ \nonumber \\
 r_{TS} &\equiv \frac{m_T}{m_S} ,~~~~
 r_\Lambda \equiv \frac{\Lambda_N}{f}, ~~~~
 r_{\text{SUSY}} \equiv \frac{m_{\text{SUSY}}}{f},
~~~ M_L= m_{\rm SUSY}~,
\label{eq:paramdef}
\end{align}
where $\zeta$ is a dimensionless parameter.
Using this parameterization, we display the constraints in the $m_{\rm
SUSY}$--$m_S$ plane. Recall that the parameter $m_{\rm SUSY}$ represents
the ``cutoff scale'' of the model while $m_S$ is the explicit
shift-symmetry breaking parameter. In Fig.~\ref{fig8100}, we have taken
$\zeta=10^{-8}$, $r_{TS}=0.1$, $r_{\Lambda}=1$, $r_{\rm
SUSY}=1$, and $\kappa = 10^{-2}$. The gray shaded region is excluded because the periodic
barrier formed when the Higgs VEV develops, cannot stop the relaxion
rolling (Eq.~\eqref{eq:StoPhiLam}). The blue shaded region is excluded
because $\phi$ never decouples from $\sigma$
(Eq.~\eqref{eq:phidecop}). In the red region, the shift-symmetry
breaking correction to the K\"ahler potential generates an inflaton mass
larger than $m_T$ (\eqref{eq:delmphi}). The green-shaded region is
disfavored since the scalar potential may become unstable in the
direction of $N\bar{N}$, as discussed in Ref.~\cite{Evans:2016htp}.
Above the dash-dotted line, $\phi_* < M_P$, and thus sub-Planckian
field values may be realized. The figure shows that supersymmetric soft
masses up to $3\times 10^5$ GeV
can be obtained for the range $10^{-10}~{\rm GeV}\lesssim m_S \lesssim
10^{-4}~{\rm GeV}$. We see that the PeV-scale SUSY region is now
constrained by the condition $\Delta m_\sigma < |m_T|$, which is a
consequence of combining the low-scale $D$-term inflation model with the
two-field relaxion model.

\begin{figure}[t]
\centering
\includegraphics[clip, width =  0.6\textwidth]{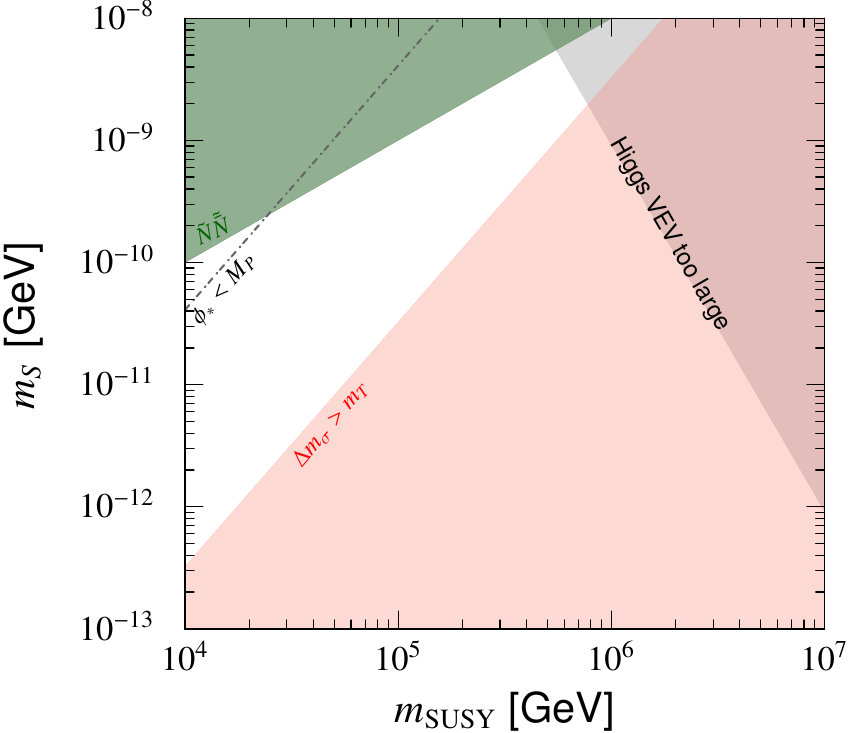}
\caption{{The allowed parameter region in the $m_{\rm SUSY}$--$m_S$
 plane, where $\zeta = 10^{-14}$, $r_{TS}=0.1$, $r_\Lambda = 1$, $r_{\rm
 SUSY} = 1$, and $\kappa = 10^{-2}$.}}
\label{fig14100}
\end{figure}

In Fig.~\ref{fig14100}, we take  $\zeta=10^{-14}$, $r_{TS}=0.1$,
$r_\Lambda=1$, $r_{\rm SUSY}=1$, and $\kappa = 10^{-2}$.  The color
coding of the excluded regions in Fig.~\ref{fig14100} is the same as in
Fig.~\ref{fig8100}. For these parameter choices the allowed region now
corresponds to supersymmetric soft mass scales up to $10^6$ GeV and
$10^{-12}~{\rm GeV}\lesssim m_S \lesssim 10^{-8}~{\rm GeV}$. It is found
that the new condition, $\Delta m_\sigma < |m_T|$ gives a very severe
limit on the parameter space in this case.

%


Finally, in the allowed region, we can always find a value of $H_I$ which satisfies the above
constraints. The lower bound on $H_I$ is given by
\begin{equation}
 H_I > \text{max}\left\{|m_S|,~4\times 10^{-9}~\text{GeV}\times
\left(\frac{m_{\rm SUSY}}{10^5~{\rm GeV}}\right)^2
\left(\frac{1}{r_{\rm SUSY}}\right)
 \right\}~,
\end{equation}
which comes from requiring that the slow roll of $\phi$, and the relaxion vacuum energy is
subdominant compared to that of the inflaton. The upper bound on the Hubble scale is
\begin{equation}
 H_I < 4.6~{\rm GeV}\times
\left(\frac{r_{TS}}{0.1}\right)^{\frac{1}{3}}
\left(\frac{1}{r_{\rm SUSY}}\right)^{\frac{1}{3}}
\left(\frac{|m_S|}{10^{-7}~{\rm GeV}}\right)^{\frac{1}{3}}
\left(\frac{m_{\rm SUSY}}{10^5~{\rm GeV}}\right)^{\frac{2}{3}}~,
\label{eq:HIrelaxbound}
\end{equation}
which comes from Eq. (\ref{eq:ClasRol}). In addition, we have upper
limits on $H_I$ to evade the cosmic string problem as discussed in
Appendix~\ref{app:cosstrourmod}.

\section{Conclusion}
\label{sec:conclusions}

In this paper, we have presented a low-scale inflationary model embedded in
a supersymmetric framework that seeks to address the hierarchy problem and
be consistent with experimental data.
Specifically, we consider a $D$-term inflationary model, characterized by a
new $U(1)$ symmetry with a FI term. There are three parameters of the
model that are relevant for the CMB phenomenology: the $U(1)$ gauge coupling,
$g$, the FI scale, $\sqrt{\xi}$, and the energy density $\rho_{\rm reh}$
at reheating (assuming an instantaneous transition between
matter domination and radiation domination).
To determine the constraints on these parameters we trade the FI scale for the
Hubble scale, $H_I$, at the moment at which the CMB modes were produced. The
measured values of the amplitude and the spectral tilt of the
primordial scalar perturbations can then be used to obtain $g$ and $H_I$ as a function of
$\rho_{\rm reh}$. By requiring $\rho_{\rm reh}^{1/4} $ to be above the
electroweak scale (in order to facilitate baryogenesis),  we find that a
value of $n_s$ compatible with the experimental limits (namely, a sufficiently
red scalar spectrum)  can be achieved provided $g \lesssim 10^{-8}$ and
$H_I \lesssim 10^5$ GeV. For this low scale of inflation, the CMB modes are
produced approximately $N_{\rm CMB} \simeq
39 \; e$-folds before the end of inflation (contrary to the 50--60
$e$-folds typically required in high scale models of inflation).
In $D$-term inflation, this relatively low value of $N_{\rm CMB}$ is used
to match the observed value of $n_s$, since deviations from scale invariance are
 inversely proportional to $N_{\rm CMB}$.

Another issue typically associated with  $D$-term inflation is the formation of cosmic
strings due to the spontaneous breaking of the $U(1)$ symmetry at the end of
inflation. We prevent this from occurring by introducing a tiny explicit breaking
of the $U(1)$ symmetry throughout the entire inflationary epoch, due to a
dynamical $D$-term mechanism. This mechanism also allows the generation of
an FI scale much below the Planck scale. Finally, a low value of $g$ is typically
problematic for reheating. For such a value, most of the energy density after
inflation is actually stored in the field that spontaneously breaks the $U(1)$ symmetry
(leading to the end of inflation). This field obtains a VEV much greater than
its mass, and therefore typically gives a large effective mass to any field that it is coupled
to with a strength greater than $g$, preventing its decay into these fields. A way to avoid
this kinematic barrier, is to introduce superpotential couplings which cancel the VEV,
so as to allow the decay into the MSSM Higgs fields, and the eventual reheating into
Standard Model fields. Thus, with the technically natural superpotential couplings and
the $U(1)$ gauge coupling $g$, a low-scale model of supersymmetric inflation can be made to
be consistent with Planck data.

This low-scale $D$-term inflation model leads to an interesting application.
It can be combined with the relaxion mechanism in order to identify the inflaton
with the second field (amplitudon) of a supersymmetric two-field relaxion model
that preserves the QCD axion solution to the strong CP problem. The inflaton
now also controls the barrier height of the relaxion periodic potential. As the
inflaton rolls, it periodically reduces the barrier height causing the relaxion to move
and scan the supersymmetric soft masses. Eventually electroweak symmetry
breaking occurs, which produces a new contribution to the relaxion barrier height
and traps the relaxion in a supersymmetry-breaking local minimum. The
correct electroweak VEV can be obtained for supersymmetric soft masses
up to the PeV scale, provided the explicit shift-symmetry breaking
parameter $m_S \lesssim 10^{-4}$~GeV, and the Hubble scale satisfies
$H_I\lesssim 10$ GeV. This dynamics takes place well
before the production of the CMB, at a time in which the energy density of the
inflaton was dominated by a quadratic (mass) term, rather than by the
Coleman--Weinberg term which instead controls the motion of the inflaton at
$N_{\rm CMB}$. The switchover between these two potential terms is a natural
consequence of the flatness associated with the logarithmic Coleman-Weinberg term,
and it distinguishes our model from other implementations of the relaxion
mechanism.  Also by identifying the amplitudon as the inflaton, a potential isocurvature
problem in the original two-field relaxion model is avoided. Therefore, the
supersymmetric inflaton-relaxion model, successfully combines low-scale $D$-term
inflation, which is technically natural, with a solution to the supersymmetric little
hierarchy problem. This intriguing connection between the
inflaton and the relaxion provides a new way to address the hierarchy problem
and deserves further study.

\section*{Acknowledgments}

We are grateful to M. Hindmarsh and A. R. Liddle for helpful correspondence, and
thank Z. Thomas for initially collaborating on the project.
The work of T.G. and M.P. is supported by the U.S. Department of Energy
Grant No. DE-SC0011842 at the University of Minnesota. The work of N.N. is
supported by JSPS KAKENHI Grant Number 17K14270.

\section*{Appendix}
\appendix

\section{Dynamical $D$-terms}
\label{app:dynamicaldterm}

\renewcommand{\theequation}{A.\arabic{equation}}
\setcounter{equation}{0}

Here, we review the dynamical generation of $D$-terms.
We basically follow the arguments in
Ref.~\cite{Domcke:2014zqa,Domcke:2017xvu}
where the dynamical generation of $D$-terms
is discussed based on the IYIT model \cite{Izawa:1996pk,
Intriligator:1996pu}. We focus on the case of the $SP(1) \cong
SU(2)$ strongly-interacting gauge theory with $N_f = 2$ quark flavors.
For more generic cases, see Ref.~\cite{Domcke:2014zqa}. In this case, we
have four chiral quark superfields $Q^i$ $(i=1,\dots, 4)$ which are in
the fundamental representation of $SP(1)$, and six singlet chiral
superfields $Z_{ij} = -Z_{ji}$ $(i,j=1,\dots, 4)$. We assign the $U(1)$
gauge charge $+1/2$ ($-1/2$) to $Q^{1,2}$  ($Q^{3,4}$), $-1$
to $Z_-\equiv Z_{12}$, $+1$ to $Z_+\equiv Z_{34}$, and $0$ to $Z_{13}$, $Z_{14}$, $Z_{23}$,
$Z_{24}$, respectively. The superpotential terms for these fundamental
fields are then given by
\begin{equation}
 W_{\rm fund} = \frac{1}{2}\sum_{i,j}\lambda_{ij} Z_{ij} Q^i Q^j ~,
\end{equation}
with $\lambda_{ij} = - \lambda_{ji}$ dimensionless Yukawa couplings. We
also couple $T$ and $\Phi_{\pm}$ to this sector via the higher-dimensional
operators $T\Phi_- Q^1 Q^2$ and $T\Phi_+ Q^3 Q^4$. In order to facilitate reheating, we will include another shift symmetric singlet, $R$, and couple it to this strongly coupled sector through the higher dimensional operator $R\Phi_+Q^1Q^2$ as well.
There are other renormalizable couplings allowed by the gauge
symmetries, such as $TZ_+ Z_-$, $Z_{\pm} \Phi_{\mp}$, $\Phi_- Q^1 Q^2$,
{\it etc.}---we simply assume that all of these unwanted terms are
negligible in the following discussion. Such a situation may be realized
by geometrically separating the $SP(1)$ sector from the
inflation/relaxion sector by means of, say, branes in extra dimensions.

Below the confinement scale of the $SP$(1) gauge interaction, $\Lambda$,
the low-energy dynamical degrees of freedom are given by the meson
fields $M^{ij} = -M^{ji} \sim Q^i Q^j/\Lambda$. The $U(1)$ charge
assignment for these meson fields follows from those for the constituent
quark fields; $M_+ \equiv M^{12}$ has $+1$, $M_- \equiv M^{34}$ has
$-1$, and the other meson fields are neutral. The meson fields are
subject to the constraint \cite{Seiberg:1994bz}
\begin{equation}
 {\rm Pf} (M^{ij}) = M^{12} M^{34} -M^{13} M^{24} + M^{14} M^{23}
=\Lambda^2 ~.
\label{eq:pfm}
\end{equation}
As in Ref.~\cite{Domcke:2014zqa}, we assume that
$\lambda_{13}$, $\lambda_{14}$, $\lambda_{23}$, and $\lambda_{24}$ are
much larger than $\lambda_+ \equiv \lambda_{12}$ and $\lambda_- \equiv
\lambda_{34}$ in order to make sure that all of the neutral fields
except $T$ remain at the origin. In this case, the condition
\eqref{eq:pfm} leads to
\begin{eqnarray}
M_+M_-=\Lambda^2 ~,
\end{eqnarray}
and the relevant part of the low-energy effective superpotential is given by
\begin{multline}
W_{\rm eff}=\kappa T \Phi_+\Phi_- + \frac{m_T}{2}T^2
+\kappa_+ T  M_+ \Phi_-
+\kappa_- T  M_- \Phi_+ \\
+\lambda_+\Lambda M_+ Z_-
+\lambda_- \Lambda M_- Z_++\kappa_1 R \Phi_+ M_- +\kappa_2 R H_u H_d +m_R R\bar R~,
\label{eq:suppot}
\end{multline}
where the third, fourth and seventh terms in the right-hand side of this
equation come from the higher-dimensional operators introduced
above.
Since these terms are generated by non-renormalizable interactions
and/or break the shift symmetry with respect to $T$ or $R$, the couplings
$\kappa_\pm$ and $\kappa_{1,2}$ can be parametrically small.

From the superpotential (\ref{eq:suppot}), we obtain the $F$-term scalar potential
as
\begin{align}
V_F
&=\Bigl|\kappa\phi_+\phi_-+
\frac{i}{\sqrt{2}}m_T \sigma
+\kappa_+ M_+ \phi_-
+\kappa_- M_- \phi_+
\Bigr|^2\nonumber \\
&+
\left|i\frac{\sigma}{\sqrt{2}}\left(\kappa \phi_-+\kappa_- M_-\right) +\kappa_1 R M_-\right|^2 + \left|i\frac{\sigma}{\sqrt{2}}\left(\kappa
 \phi_++\kappa_+ M_+\right) \right|^2
 \nonumber \\
&+\Bigl|\frac{i\kappa_+}{\sqrt{2}} \sigma \phi_-
+\lambda_+ \Lambda Z_-\Bigr|^2
+\Bigl|\frac{i\kappa_-}{\sqrt{2}} \sigma \phi_+
+\lambda_- \Lambda Z_++\kappa_1 R \phi_+\Bigr|^2
\nonumber \\ &+\Bigl| \kappa_1 \phi_+M_- +\kappa_2 H_u H_d +m_R \bar R\Bigr|^2+\left| m_R R\right|^2 \nonumber  \\
& +\left|\lambda_+ \Lambda M_+\right|^2+\left|\lambda_- \Lambda M_-
\right|^2  ~,
\label{eq:vfindymod}
\end{align}
where we have assumed $|\sigma| \gg |\tau|$ as in Section~\ref{sec:dtermmodel}.
There is also a $D$-term contribution to the scalar potential
\begin{equation}
V_D=\frac{g^2}{2}\left(|\phi_+|^2-|\phi_-|^2+|Z_+|^2-|Z_-|^2
+|M_+|^2-|M_-|^2-\xi_{\rm tree}\right)^2~,
\end{equation}
where $\xi_{\rm tree}$ denotes the tree-level FI term, which can be
taken to be zero when the dynamical sector generates a large enough
contribution for $D$-term inflation to work. This amounts to the
difference of the VEVs of $M_\pm$ being large enough.

Now to leading order\footnote{We have checked this perturbatively in the
limit where $g$, $\kappa_\pm$, and $\kappa_1$ are small.} in
$\kappa_\pm$, $\kappa_1$ and $g$,
the $F$-terms vanish in the vacuum except for $Z_\pm$ and $T$ with the
fields having the following VEVs,\footnote{The leading order
contributions to the VEVs of all fields except $\phi_+$ and $M_\pm$ can be
taken to be zero in the limit $\kappa_-\to 0$. This has no adverse
effect on the model we consider.  }
\begin{align}
\langle M_\pm \rangle &=
\sqrt{\frac{\lambda_\mp}{\lambda_\pm}}\Lambda ~, \label{eq:mpmvev}\\
\langle \phi_{\pm}\rangle &=-\frac{\kappa_\pm}{\kappa}
\sqrt{\frac{\lambda_\mp}{\lambda_\pm}}\Lambda ~, \label{eq:phipm}\\
\langle Z_\pm \rangle &=
\frac{i\kappa_+\kappa_-}{\kappa \sqrt{2 \lambda_+ \lambda_-}} \sigma~, \\
\langle R \rangle &=-i\frac{\kappa_1\kappa_- \lambda_+ \sigma \Lambda^2}{\kappa_1^2\lambda_+ \Lambda^2+\lambda_- m_R^2}~,
\end{align}
and $\bar R$ can be found by solving $F_R$. We have assumed $g \ll \lambda_{\pm}$ as in
Ref.~\cite{Domcke:2014zqa}. This condition can easily be satisfied in
the case of low-scale $D$-term inflation as can be seen from
Eq.~\eqref{eq:gu1}. The details of the calculation for the VEVs of
$M_\pm$ can be found in Ref.~\cite{Domcke:2014zqa}. By using
Eq.~\eqref{eq:mpmvev}, we then obtain the dynamically generated FI term:
\begin{equation}
 \xi_{\rm dyn} = \left(\frac{\lambda_+}{\lambda_-} - \frac{\lambda_-}
{\lambda_+}\right) \Lambda^2 ~.
\end{equation}
This can explain the required value shown in Eq.~\eqref{eq:xisq} if
$\Lambda \simeq 10^{16}$~GeV. Such a dynamically-generated FI term has
several advantages. First, this can naturally explain why the FI
term is much smaller than the fundamental scale, such as the Planck
scale. In addition, this allows the model to couple with supergravity in
a consistent manner, which is very difficult if the theory possesses a
constant FI term.\footnote{As pointed out in
Refs.~\cite{Komargodski:2009pc, Komargodski:2010rb}, the Ferrara--Zumino
current multiplet \cite{Ferrara:1974pz}, which contains the
energy-momentum tensor and the supersymmetry current, becomes
gauge-variant in the presence of a constant FI term, and thus cannot be
well-defined. This prevents the theory from coupling to minimal
supergravity. } It also provides a means to suppress cosmic strings and facilitate reheating through additional couplings of the dynamical sector to the inflaton sector.

As mentioned above, the fields $Z_\pm$ develop $F$-terms
\begin{equation}
 F_{Z_\pm} = - \sqrt{\lambda_+ \lambda_-} \Lambda^2 ~.
\end{equation}
Note that since we have assumed $g\ll \lambda_{\pm}$, this $F$-term VEV
is much larger than the dynamically generated $D$-term $g\xi_{\rm dyn}
\sim g \Lambda^2$. The size of $F_{Z_\pm}$ can, however, be much smaller
than $\Lambda^2$ if one takes $\lambda_\pm$ to be very small.

Since this setup introduces another source of SUSY-breaking as well as
the shift-symmetry breaking, this sector may give rise to a sizable
shift-symmetry breaking effect on the inflaton/amplitudon field. For
example, if $F_{Z_\pm}$ induces the gravity-mediated soft masses of
$\phi_\pm$, this generates a mass for $\sigma$ via the Coleman-Weinberg
potential as in Eq.~\eqref{eq:delmphicw} of order
\begin{equation}
 \Delta m_\sigma \simeq \frac{\kappa}{4\pi} \frac{|F_{Z_\pm}|}{M_P} ~,
\end{equation}
and thus the requirement \footnote{Note, this constraint is only important if we wish to identify the inflaton as the amplitudon. However, there is still a restriction on the couplings $\lambda_\pm$ but it is much weaker.}
of $\Delta m_\sigma < m_T$ restricts
$\sqrt{\lambda_+ \lambda_-}$ as
\begin{equation}
 \sqrt{\lambda_+ \lambda_-} < 3\times 10^{-18} \times \left(\frac{|m_T|}{10^{-7} ~{\rm
			       GeV}}\right)
\left(\frac{\kappa}{10^{-2}}\right)^{-1}
\left(\frac{\Lambda}{10^{16}~{\rm GeV}}\right)^{-2}
~.
\end{equation}
This may be in contradiction with the condition $g\ll \lambda_{\pm}$. However,
if we consider the no-scale K\"{a}hler terms for $\phi_\pm$, the
gravity-mediated mass terms may vanish and thus the dominant
contribution comes from anomaly mediation \cite{Randall:1998uk,
Giudice:1998xp}. In this case, $\Delta m_\sigma$ is suppressed by
another factor of $\kappa^2/(16\pi^2)$, which allows $\lambda_\pm$ to
be larger than the gauge coupling constant.

If ${Z_\pm}$ were to couple to $T$ too strongly, their $F$-terms,
$F_{Z_\pm}$, would generate a large mass for $\sigma$. However, $Z$ does
not interact with $T$ even at one-loop level, and thus $F_{Z_\pm}$ do
not give sizable effects on $\sigma$ through radiative corrections. The $F$-terms for $M_\pm$ vanish at the leading order with
respect to the small couplings $g$, $\lambda_\pm$, and $\kappa_\pm$, and
thus their effects are very tiny and can be completely neglected. On the
other hand, there are non-zero contributions to the $F$-terms of
$\phi_\pm$, which can induce a mass term for the $\sigma$ field through
the terms in the second line of Eq.~\eqref{eq:vfindymod}. From a straightforward
calculation, we find that this effect is smaller than the
Coleman--Weinberg effect if
\begin{equation}
 \left[
|\kappa_+M_+|^2+
|\kappa_-M_-|^2
\right]^{\frac{1}{2}}
<
\frac{\kappa g \sqrt{\xi}}{4\pi} ~.
\label{eq:kappauplim}
\end{equation}
Because $\langle M_\pm \rangle \sim \sqrt{\xi}$, this roughly places a constraint of $\sqrt{|\kappa_+|^2+|\kappa_-|^2} \lesssim \frac{\kappa g}{4\pi}$. Because $g$ grows with the inflation scale, this constraint on $\kappa_\pm$ becomes weaker for larger inflation scales. Thus, we find that there is a sufficient range of parameter space where both
\eqref{eq:kaplowlim} and \eqref{eq:kappauplim} are satisfied.

\section{Cosmic Strings and Inflation}
\label{app:inflation}

\renewcommand{\theequation}{B.\arabic{equation}}
\setcounter{equation}{0}

In this section, we discuss the effects of quantum and thermal
corrections on the $U(1)$ symmetry breaking during inflation.
This breaking can prevent the generation of cosmic strings after inflation.

\subsection{Quantum Fluctuations and Cosmic Strings}

\subsubsection{General Model of Cosmic Strings}

First, we discuss the effects of quantum fluctuations on the
$U(1)$-breaking scalar field during inflation. To that end, we consider a
simple toy model which is described by the following Lagrangian:
\begin{equation}
 {\cal L} = |\partial_\mu \phi|^2 - V(\phi) ~,
\label{eq:simplemodel}
\end{equation}
with
\begin{equation}
V(\phi)=\frac{\bar{g}^2}{2}\left(|\phi|^2-\bar{\xi}\right)^2-\phi C -\phi^*
 C^* +|\bar{\kappa}|^2|I|^2|\phi|^2 ~,
\label{eq:simplemodelpot}
\end{equation}
where $I$ is the inflaton. In the limit that $C=0$, the theory has a
global $U(1)$ symmetry, and the vacuum
corresponds  to $\phi=0$ for $|\bar{\kappa} I|^2 > \bar{g}^2\bar{\xi}$.
Instead, when $|\bar{\kappa} I|^2 <\bar{g}^2\bar{\xi}$ we obtain
\begin{equation}
|\phi|^2 =\bar{\xi} ~.
\label{eq:vacbrok}
\end{equation}
In this case, the global $U(1)$ symmetry is spontaneously broken,
and the vacuum manifold is $U(1)\cong S^1$ as seen in Eq.~\eqref{eq:vacbrok}.
Since the first homotopy
group of this manifold is $\pi_1 (U(1)) \cong \mathbb{Z}$, vortices and
strings can form in three and four spacetime dimensions, respectively.
Here, we consider four spacetime dimensions and take the $z$ axis
parallel to the string. We use polar coordinates $(r, \theta)$ in the
$x$--$y$ plane, with the origin located at the center of the cosmic
string.\footnote{The translational invariance in the theory is also
spontaneously broken. }

Let us explicitly see how strings form after the $U(1)$ symmetry
is broken. In order for the energy per unit length of the string, or
string tension, to be finite,
it is necessary that $|\phi| \to \sqrt{\bar{\xi}}$ as $r \to \infty$.
However, the phase of $\phi$ at infinity is not necessarily the same in
different directions; for instance, we may have
\begin{equation}
 \phi (x) \to \sqrt{\bar{\xi}} e^{in\theta} ~~~~~~
(r \to \infty) ~,
\end{equation}
with $n$ an integer. Such non-trivial field configurations (for
$n\neq 0$) correspond to the formation of strings in the system.

So far, we have considered global strings. It turns out, however, that
the string tension in this case diverges if the spatial volume is
infinite. On the other hand, if the $U(1)$ symmetry is a gauge symmetry,
then the tension becomes finite thanks to non-trivial field configurations of
the $U(1)$ gauge field. In this case, the winding number $n$ corresponds
to the magnetic flux in the string core. If the $U(1)$ charge times the
gauge coupling of $\phi$ is given by $\bar{g}$, then the masses of the
$U(1)$ gauge boson and $\phi$ in the broken phase are equal; in this
case, we have BPS strings whose string tension is given by
\begin{equation}
 {\mu} = 2\pi \bar{\xi} n ~.
\end{equation}
The model discussed in Section \ref{sec:cosmic-string}
assumes a $U(1)$ gauge symmetry which
gives rise to BPS strings. The behavior of the $U(1)$ symmetry
breaking itself can, however, be captured with the simplifed model in
Eq.~\eqref{eq:simplemodel}, and thus we focus on this in what follows.


\subsubsection{Effects of the Linear Term}
\label{app:effoflin}

Here we examine in detail what happens if $C\ne 0$, which explicitly
breaks the $U(1)$ symmetry. To that end, let us take
\begin{equation}
 \phi = {v_r} e^{i\alpha} ~.
\end{equation}
Then, Eq.~\eqref{eq:simplemodelpot} leads to
\begin{equation}
 V (\phi) = \frac{\bar{g}^2}{2}\left(v_r^2 -\bar{\xi}\right)^2
-v_r e^{i\alpha} C - v_r e^{-i\alpha} C^*
+|\bar{\kappa}|^2 |I|^2 v_r^2~,
\label{eq:vphipolar}
\end{equation}
with the vacuum conditions
\begin{align}
  2 \bar{g}^2 v_r \left(v_r^2 -\bar{\xi}\right) +
2|\bar{\kappa}|^2 |I|^2 v_r &=
e^{i\alpha} C + e^{-i\alpha} C^* ~,
\label{eq:vevcondvr}
\\
 e^{2i\alpha} &= \frac{C^*}{C} ~.
\label{eq:vacalp}
\end{align}
From Eq.~\eqref{eq:vacalp}, we see that the phase of $\langle \phi
\rangle$ is uniquely selected by the phase of the $U(1)$ symmetry
breaking term such that
\begin{equation}
 \alpha \equiv - {\rm arg}(C) ~~~~~~ ({\rm mod}~ 2\pi) ~,
\label{eq:phaseofvev}
\end{equation}
and therefore strings never form. We use this mechanism to evade the formation of cosmic strings.

With the condition \eqref{eq:phaseofvev}, Eq.~\eqref{eq:vevcondvr} leads
to
\begin{equation}
   2 \bar{g}^2 v_r \left(v_r^2 -\bar{\xi}\right) +
2|\bar{\kappa}|^2 |I|^2 v_r = 2|C| ~.
\end{equation}
Assuming that $\bar{g}$ is very small so that we can neglect the first
term, as justified in our $D$-term inflation model, we obtain $v_r$
as
\begin{equation}
 v_r \simeq \frac{|C|}{|\bar{\kappa} I|^2} ~.
\label{eq:vrapprx}
\end{equation}
In order for strings not to form, we require that this VEV is larger
than quantum fluctuations in $v_r$ induced by inflation and thermal
fluctuations. This would guarantee that all patches of the sky will have
the same phase of $\langle \phi \rangle$ in the end,  so no
strings could form.

\subsubsection{Cosmic Strings For Our Model}
\label{app:cosstrourmod}

If the quantum fluctuations around the time inflation ends are large
enough, cosmic strings could still form since there could be
fluctuations of $\phi$ which spoil the phase alignment imposed by the
$U(1)$-breaking term.\footnote{This
may not be a necessary condition for no strings, but it is
sufficient. As long as there is only one minimum of the potential for
some time after the CMB is set, the quantum fluctuations will no longer
be correlated on superhorizon scales and large strings will not
form. Smaller strings may still form, but they lead to different
problems, such as overclosure,
and are not constrained by the CMB.} Here we evaluate the size
of the quantum fluctuations in $\alpha$ from inflation; if the
fluctuation $\delta\alpha$ can be as large as $\pi$, then strings can
form after inflation. During inflation, the size of the
fluctuations depends on the size of the fields mass in the $\alpha$
direction, $m_\alpha$, relative to the Hubble parameter during
inflation, $H_I$. The mass $m_\alpha$ can be obtained from
Eq.~\eqref{eq:vphipolar} as\footnote{To obtain a mass term for the
canonically normalized field, we need to rescale by $\sqrt{2} v_r$. }
\begin{align}
m_\alpha^2 = \frac{1}{2v_r^2}\frac{\partial^2 V}{\partial \alpha^2}
 \biggr|_{\alpha =-{\rm arg}(C)} = \frac{|C|}{v_r} \simeq
|\bar{\kappa} I|^2 ~,
\label{eq:malpha}
\end{align}
where we have used Eq.~\eqref{eq:vrapprx}.
Since the cosmic strings that can be see in the CMB have lengths that
are of order the current horizon size, the variations in the phases must
be in place when the inflaton has the field value corresponding to
$N_{\rm CMB}$. The fluctuations in the $\alpha$ direction at this time
is then estimated as
\begin{equation}
 \langle \delta\alpha^2 \rangle = \frac{H_I^3}{12\pi^2 m_\alpha v_r^2}
\simeq \frac{H_I^3 |\bar{\kappa} I_{\rm CMB}|^3}{12 \pi^2 |C|^2 } ~,
\end{equation}
where $I_{\rm CMB}$ denotes the field value of $I$ at the time when the
CMB is set. The condition $\sqrt{\langle \delta\alpha^2 \rangle} < \pi$
imposes a lower bound on $|C|$, given an inflation model.


We now apply this result to the model discussed in
Appendix~\ref{app:dynamicaldterm}. This model can be mapped on to the
simplified model we considered above by setting $\bar{g} = g$,
$\bar{\xi} = \xi$, $\bar{\kappa} = \kappa$,  $\phi = \phi_+$, $I=
i\sigma/\sqrt{2}$, and $C= \kappa \kappa_+^* M_+^* \sigma^2/2$.
Note that from Eqs.~\eqref{eq:SigCMB}, \eqref{eq:spectilt},
\eqref{eq:PowSpec}, \eqref{eq:xisq}, and \eqref{eq:gu1}, we have
$\sigma_{\rm CMB} = H_I  /(\pi A_s^{1/2}(1-n_s))$. The limit
$\sqrt{\langle \delta\alpha^2 \rangle} < \pi$ then reads
\begin{align}
 H_I &<
\left[\frac{12\pi^3 |\kappa_+|^2 |M_+|^2 }{\sqrt{2}\kappa
	A_s^{\frac{1}{2}} (1-n_s)}\right]^{\frac{1}{2}}~.
	\label{eq:CMBStrings}
\end{align}
We can easily find a set of parameters which satisfy this condition as
well as Eq.~\eqref{eq:kappauplim}.

Since the mass of $\phi_+$ approaches zero at the end of inflation, it
is possible that strings with size much smaller than the current horizon
could have formed if the fluctuations during this period are too
large. To verify that this is not a problem, we examine the same
constraint but in the case that $m_\alpha=0$, which leads to
fluctuations of order $\langle \delta\alpha^2 \rangle =H_I^2/(4\pi^2
v_r^2)$. In this case the constraint becomes
\begin{align}
 H_I <
2\pi^2 \frac{ |\kappa_+| }{\kappa}|M_+|
=2\times 10^7~{\rm GeV} \times
\left(\frac{|\kappa_+|}{10^{-12}}\right)
\left(\frac{|M_+|}{10^{16} ~{\rm GeV}}\right)
\left(\frac{\kappa}{10^{-2}}\right)^{-1}~.
\label{eq:Hmass0}
\end{align}
This is again compatible with the condition \eqref{eq:kappauplim}. Since the constraints on strings with sizes much smaller than the CMB scale may not be problematic, we only cite the constraint in Eq. (\ref{eq:CMBStrings}) in the main text.  However, since (\ref{eq:Hmass0}) scales with $\kappa_+$, just like the constraint in Eq. (\ref{eq:CMBStrings}), this constraint can be satisfied for any $H_I$ by choosing an appropriate $\kappa_+$ that is consistent with Eq. (\ref{eq:kappauplim}).


\subsection{Thermal Fluctuations After Inflation}
\label{app:thermfluc}

Finally, we need to consider the effect that thermal
fluctuations can have on the formation of cosmic strings. Since we are
only interested in an order of magnitude estimate, we will use
\begin{equation}
\langle \delta\alpha ^2 \rangle_{\rm therm} \simeq T^2/v_r^2  ~,
\label{eq:ThermFluc}
\end{equation}
where the fluctuations are probably a bit more mild than this. By
requiring $\sqrt{\langle \delta\alpha ^2 \rangle}_{\rm therm} < \pi$,
we then obtain an upper bound on the reheating temperature as $T_R
< \pi v_r$. This corresponds to $T_R < \pi |\langle \phi_+ \rangle|$ in
the $D$-term inflation model.

During most of inflation, the non-zero VEV for $\phi_+$ is determined by
the linear term and the mass term as we have seen above. However,
towards the end of inflation,
\begin{eqnarray}
|\kappa T_c|^2=\frac{\kappa^2}{2}\sigma_c^2 = g^2\xi~,
\label{eq:U1Break}
\end{eqnarray}
and so the $\phi_+$ mass approaches zero.  When Eq.~\eqref{eq:U1Break}
is satisfied, the mass term is zero, and the VEV is set by the quartic
term and the linear term, which can be read from Eq.~\eqref{eq:vevcondvr}
as
\begin{equation}
 |\langle \phi_+ \rangle| = \biggl|\frac{\kappa_+
  M_+ \xi}{\kappa}\biggr|^{\frac{1}{3}} ~.
\end{equation}
Thus, the upper bound on $T_R$ is given by
\begin{align}
 T_R &< \pi \biggl|\frac{\kappa_+ M_+ \xi}{\kappa}\biggr|^{\frac{1}{3}}
\nonumber \\
&=
6.5 \times 10^{14}~{\rm GeV} \times
\biggl(\frac{|\kappa_+|}{10^{-12}}\biggr)^{\frac{1}{3}}
\biggl(\frac{|M_+|}{10^{16}~{\rm GeV}}\biggr)^{\frac{1}{3}}
\biggl(\frac{\kappa}{10^{-2}}\biggr)^{-\frac{1}{3}}
\biggl(\frac{1-n_s}{0.03}\biggr)^{\frac{1}{6}}
\biggl(\frac{A_s}{2.1\times 10^{-9}}\biggr)^{\frac{1}{6}}
~,
\end{align}
where we have used Eq.~\eqref{eq:xisq}. For the values of $H_I$ we are
considering, the reheating temperature can easily satisfy this
constraint. However, above we have assumed that the field is not displaced from its minimum as the location of the minimum moves near the end of inflation. If the field can indeed track its minimum, there will be no cosmic strings from thermal
fluctuations.


Now we need to verify that $\phi_+$ can track the minimum as the
inflaton approaches the critical value, or at least track it
sufficiently long that thermal fluctuations do not cause strings to
form. The minimum begins to move once $|\kappa \sigma|^2 \sim
g^2\xi$. To approximate the actual size of the VEV of $\phi_+$ once
$\sigma$ hits its critical value, we estimate how far $\phi_+$ will
track its minimum.  The field $\phi_+$ will track its minimum until its mass
$m_{\phi_+}$ becomes equal to $3H_I/2$, a well-known
relationship. During inflation, the VEV of $\phi_+$ has plenty of time
to settle into its minimum of $|\kappa_+ M_+/\kappa|$, where the mass of
$\phi_+$ (both the radial and phase directions) is given by
$m_{\phi_+}^2 = \frac{\kappa^2 \sigma^2}{2} - g^2 \xi$. The value of
$\sigma$ where $\phi_+$ ceases to track its minimum can then be
found from $m_{\phi_+} = 3H_I/2$. At this point, the $\phi_+$ VEV is (see
Eq.~\eqref{eq:vrapprx}) given by
\begin{align}
| \langle \phi_+ \rangle | \simeq \frac{|C| }{m_{\phi_+}^2}
= \frac{2|\kappa \kappa_+ M_+ \sigma^2|}{9 H_I^2}
\simeq \frac{4g^2 \xi|\kappa_+ M_+ |}{9 \kappa H_I^2}
\simeq
\frac{8}{3\sqrt{3 A_s (1-n_s)}} \biggl|\frac{\kappa_+ M_+ }{\kappa}
\biggr|
~,
\end{align}
where we have used $\frac{\kappa^2 \sigma^2}{2} \simeq g^2 \xi $,
Eq.~\eqref{eq:PowSpec}, and Eq.~\eqref{eq:hiandgxi}. The upper bound on
$T_R$ is then given by
\begin{align}
 T_R &< \frac{8 \pi}{3\sqrt{3 A_s (1-n_s)}} \biggl|\frac{\kappa_+ M_+ }{\kappa}
\biggr| \nonumber \\
&=
1.5\times 10^{19} ~{\rm GeV}\times
\biggl(\frac{|\kappa_+|}{10^{-12}}\biggr)
\biggl(\frac{|M_+|}{10^{16}~{\rm GeV}}\biggr)
\biggl(\frac{\kappa}{10^{-2}}\biggr)^{-1}
\biggl(\frac{1-n_s}{0.03}\biggr)^{-\frac{1}{2}}
\biggl(\frac{A_s}{2.1\times 10^{-9}}\biggr)^{-\frac{1}{2}}
~.
\end{align}
This bound is satisfied for the entire parameter space that is
compatible with the CMB observation. Indeed, we can show that the
maximum possible reheat temperature is always smaller than the VEV of
$\phi_+$.

{\small
\bibliographystyle{JHEP}
\bibliography{ref}
}

\end{document}